\documentclass[conference]{IEEEtran}
\IEEEoverridecommandlockouts
\usepackage{cite}
\usepackage{amsmath,amssymb,amsfonts}
\usepackage{graphicx}
\usepackage{textcomp}

\usepackage{graphicx}  
\usepackage{float}     
\usepackage{booktabs}  
\usepackage{caption}   
\usepackage{subcaption} 
\usepackage{float}
\usepackage{hyperref}  
\hypersetup{
    colorlinks=true,
    linkcolor=blue,
    citecolor=blue,
    urlcolor=blue
}
\usepackage{amsmath}
\usepackage{algorithm}
\usepackage{algpseudocode}
\usepackage{textcomp}
\usepackage{array}
\usepackage[table,xcdraw]{xcolor}
\usepackage{tabularx}
\newcolumntype{Y}{>{\centering\arraybackslash}X}

\usepackage{booktabs}
\usepackage{array}
\newcolumntype{P}[1]{>{\centering\arraybackslash}p{#1}}
\usepackage{stfloats}
\usepackage{placeins}

\def\BibTeX{{\rm B\kern-.05em{\sc i\kern-.025em b}\kern-.08em
    T\kern-.1667em\lower.7ex\hbox{E}\kern-.125emX}}
\begin{document}

\title{Improving the Energy-Efficiency of the Code Generated by LLMs through Effective Prompting}

\author{
\IEEEauthorblockN{
\textbf{Ritika Rekhi}\IEEEauthorrefmark{2},
\textbf{Bing Zhang}\IEEEauthorrefmark{3},
\textbf{Md Arman Islam}\IEEEauthorrefmark{2},
\textbf{Jaya Krishna Pasham}\IEEEauthorrefmark{2},
\textbf{Yeswanth Chitturi}\IEEEauthorrefmark{2},
}
\IEEEauthorblockN{
\textbf{Akshay Paramesha}\IEEEauthorrefmark{2},
\textbf{Isha Valiveti}\IEEEauthorrefmark{2},
\textbf{Asif Imran}\IEEEauthorrefmark{2},
\textbf{Bekir Turkkan}\IEEEauthorrefmark{4},
\textbf{Tevfik Kosar}\IEEEauthorrefmark{2}
}
\IEEEauthorblockA{\IEEEauthorrefmark{2}University at Buffalo, Buffalo, NY, USA}
\IEEEauthorblockA{\IEEEauthorrefmark{3} National Center for Supercomputing Applications, Urbana, IL, USA}
\IEEEauthorblockA{\IEEEauthorrefmark{4}IBM Research, Yorktown Heights, NY, USA}
}

\maketitle

\begin{abstract}
As AI-assisted programming becomes increasingly mainstream, the environmental impact of AI-generated software has emerged as an important consideration. This motivates evaluating LLM-generated code beyond functional correctness by considering execution efficiency and energy consumption. However, despite substantial advances in code generation, frontier LLMs are rarely evaluated based on the energy efficiency of the code they produce. In this work, we conduct a comprehensive evaluation of 21 prompting strategies for energy-efficient code generation and identify 8 strategies for evaluation across 10 widely used open-weight and proprietary LLMs. We evaluate their effectiveness for both Python and C++ code generation relative to a baseline prompt. Across the evaluated models, the selected prompting strategies achieved energy reductions of up to 25\% for Python and 17\% for C++ code generation. At the model level, Python energy reductions reached up to 50\% for \mbox{Granite-4.0-H-Small}, 39\% for \mbox{Claude 4.5 Haiku}, and 28\% for \mbox{MiniMax M3}, while C++ reductions reached up to 57\% for \mbox{Granite-4.0-H-Small} and 6.4\% for \mbox{Qwen3-Coder-480B-A35B-Instruct}. These results demonstrate that prompting strategies can substantially influence the energy consumption of LLM-generated code, although their effectiveness varies across models and programming languages. Our findings highlight the importance of incorporating energy efficiency into the evaluation and optimization of LLM-based code generation and provide practical insights into designing prompts for more sustainable AI-assisted programming.
\end{abstract}


\section{Introduction}
The growing adoption of Large Language Models (LLMs) in software development, along with their increasing role in the day-to-day workflows of developers, has created new challenges regarding the amount of energy required to execute code produced by LLMs. While earlier generations of coding assistants were able to assist developers with only a single line of code at a time, today's AI coding assistants complete entire projects with minimal human input. As a result, AI-assisted development has become a common practice, with 84\% of surveyed developers using or planning to use AI tools and 51\% of professional developers reporting daily use~\cite{stackoverflow2025ai}.

A recent analysis of more than 30 million GitHub commits by around 160,000 developers found that AI wrote roughly 29\% of the Python functions contributed by U.S.\ developers by the end of 2024~\cite{daniotti2026}, with developers in 2026 reporting that about 51.9\% of their code was AI-generated~\cite{dx2026}. The energy costs associated with AI-created code are quite high. In 2025, the world's data centers consumed roughly 485 TWh of electricity, a 17\% increase over the previous year. This represents around 1.5\% of all of the world's electrical needs. However, it is expected to grow to nearly twice that size by 2030, reaching 950 TWh~\cite{iea2026keyquestions}. In fact, U.S.\ based research estimates that data centers could consume about 649 TWh by 2030, accounting for 11.8\% of total U.S. electricity, and as much as 15.3\% under higher-demand estimates~\cite{lbnl2025datacenter}. Additionally, it is estimated that globally, the IT sector is responsible for roughly 1.5\% to 4\% of global greenhouse gas emissions~\cite{bieser2023review,freitag2021climate}, comparable to the emissions of the airline industry worldwide~\cite{atag2023facts}.

The energy efficiency of LLM-generated code is often much lower than that of human-written solutions, and in some cases LLM-generated code can consume up to 450 times more energy than human-generated canonical solutions~\cite{ourpreviouswork}.
One key reason is that LLMs frequently use heavier algorithms, additional data structures, and recursive logic, which involve unnecessary computations.
This is not a one-time cost, as the energy consumed by code recurs every time it runs. This runtime energy adds to the electricity demand of data centers and the load on already strained power grids. Hence, making LLM-generated code more efficient matters as much for sustainability as it does for performance.

Larger models and higher pass rates did not correlate with more efficient code execution~\cite{niu2024efficiency}. Given the high cost of re-training a pretrained model for energy awareness, prompting provides 
an alternative method. It is applicable to any model and incurs virtually no cost. However, despite considerable research effort, the current body of evidence on
whether prompts reduce energy expenditure is mixed, and
nearly all evidence has focused solely on Python~\cite{cappendijk2024generating, ashraf2025greencode, apsan2026generating}.

To investigate whether prompt design can improve the energy efficiency of LLM-generated code and whether these improvements generalize across programming languages and models, we first evaluated 21 prompting techniques on Python code generated by 3 LLMs using a stratified subset of benchmark problems. Based on these results, we selected 8 promising prompting strategies and evaluated them across 10 popular open-weight and proprietary LLMs for both Python and C++. Our evaluation used 878 benchmark problems from EffiBench~\cite{huang2024effibench}, which we extended with an execution and validation pipeline for C++ code. 


In summary, the paper makes the following contributions:
\begin{enumerate}

\item We conduct a comprehensive evaluation of 21 prompt engineering strategies for energy-efficient code generation using a stratified subset of 110 problems from the extended EffiBench dataset. The evaluated prompts span both single-turn and multi-turn interactions and combine role-based prompting, few-shot learning, and chain-of-thought reasoning with targeted modifications, including explicit energy efficiency instructions, coding guidelines, and energy-oriented code examples. This systematic design enables us to isolate and identify prompt characteristics that most effectively improve the energy efficiency of LLM-generated code.

\item From our initial analysis, we identify 8 prompting strategies that consistently improve the energy efficiency of generated code and evaluate them using 878 problems across 10 popular open-weight and proprietary LLMs. 

\item We develop a reproducible methodology for accurately measuring the energy consumption of LLM-generated code. Our evaluation framework incorporates language-specific execution environments, standardized benchmarking procedures, and carefully designed baseline and cool-off intervals to minimize measurement noise and enable fair comparisons of prompting strategies.

\end{enumerate}

\section{Related Work}

{\bf Green Software Practices:}
Energy costs associated with software have been of concern prior to Large Language Models (LLMs). Lago et al.~\cite{lago2015framing} point out that sustainability should be considered a primary software quality property rather than an afterthought, and Pinto and Castor~\cite{pinto2017energy} add that energy efficiency is a major issue for application developers and not just for hardware or data center managers. 
Imran et al.~\cite{imran2024towards} proposed systematic techniques to identify energy leaks in Java and Python code.
Kumar et al.~\cite{kumar2026energy} analyzed the impact of dataframe libraries on the energy consumption of the code. 
Apsan et al.~\cite{apsan2026generating} gathered 28 guidelines for writing energy-efficient Python, organized into families such as code optimization, multithreading, native code, function calls, object orientation, and network use. Pereira et al.~\cite{pereira2017energy} measured the energy consumption of 27 programming languages, concluding that the fastest language may not be the most energy-efficient. 

Regarding measurements of energy usage, Henderson et al. ~\cite{henderson2020towards} proposed a standardized methodology to measure and report both the energy and carbon emissions associated with computational experimentation. In addition, Lannelongue et al.~\cite{lannelongue2021green} developed the Green Algorithm technique to estimate the equivalent carbon emissions of runtime and hardware configurations. As interest in energy usage has grown, so too has the focus on the total cost of AI, beginning with Strubell et al.~\cite{strubell2019energy}, who estimated the energy and carbon usage required to train NLP models; followed by calls for ``Green AI'' that would include reporting energy-efficiency as part of normal evaluation metrics, rather than having it be an option~\cite{schwartz2020green}.

{\bf Efficiency of LLM-generated Code:}
Increasing numbers of benchmarks demonstrate that although LLMs generate functionally correct code, the generated code is often inefficient. The EffiBench project~\cite{huang2024effibench} paired each test problem with a hand-coded optimized version, demonstrating that on average the LLM output was much slower. Mercury~\cite{du2024mercury} expanded upon this, providing a dual-metric to assess both functional correctness and efficiency of generated code. Moreover, Niu et al.~\cite{niu2024efficiency} separated two previously conflated aspects of evaluating generated code: correctness and runtime efficiency. Their findings indicated that neither achieving a higher pass rate nor increasing model size resulted in generating code that ran faster. The majority of prior research focused on assessing the runtime of generated code. Our previous research assessed the energy use of 20 different LLMs on 878 different problems~\cite{ourpreviouswork}.
We observed that LLM-generated code was generally less energy-efficient than human-written solutions, with differences reaching several orders of magnitude for certain problem types. Similar findings have been reported in prior studies across different settings, including comparisons of the energy consumption of small and large language models~\cite{ashraf2025energyaware}, as well as comparisons between LLM-generated code and solutions developed by expert green-software developers across multiple hardware platforms~\cite{apsan2026generating}.

{\bf Prompting for Efficient Code Generation:} The substantial computational and financial costs of retraining LLMs have motivated prompt-based approaches for improving code efficiency at inference time without modifying the underlying model. Niu et al.~\cite{niu2024efficiency} showed that explicitly prompting models to optimize code, particularly when incorporating a reasoning step, can yield modest but measurable improvements in execution time. Apsan et al.~\cite{apsan2026generating} showed that different prompting techniques can have varying and inconsistent effects on the energy efficiency of LLM-generated Python code, with the most energy-efficient prompting strategy varying across execution environments. Cappendijk et al.~\cite{cappendijk2024generating} reported energy reductions of up to 59.7\% for certain problems, but also found that not all problems benefited from prompting and that one prompt increased energy consumption by more than fourfold. Similarly, Ashraf et al.~\cite{ashraf2025greencode} found that chain-of-thought prompting most consistently reduced energy consumption for smaller models, whereas few-shot prompting produced such improvements less frequently, with the optimal prompting strategy depending on the model. Jonnala et al.~\cite{jonnala2025measuring} evaluated the impact of Chain-of-Thought (CoT) prompting on the efficiency of LLM-generated Python code using Greedy problems from the EffiBench dataset and their results showed that the impact of CoT prompting on code efficiency varied across the evaluated models and efficiency metrics. Tuttle et al.~\cite{tuttle2024generating} investigated the energy efficiency of code generated by eight LLMs on LeetCode problems, finding that LLM-generated solutions generally lagged behind the most efficient solutions and that the effectiveness of prompting strategies varied across models. Cursaru et al.~\cite{cursaru2024controlled} evaluated the energy efficiency of Code Llama-generated code across C++, JavaScript, and Python, finding that human-written implementations were generally more energy-efficient, while explicitly prompting the model for energy-efficient code resulted in comparable or worse energy efficiency depending on the programming language and problem. Peng et al.~\cite{peng2025perfcodegen} introduced \textit{PerfCodeGen}, a framework designed to improve the correctness and runtime performance of LLM-generated code. Their study also found that certain prompting techniques improved performance relative to the baseline prompt, although they did not consistently outperform \textit{PerfCodeGen}. Huang et al.~\cite{huang2024effilearner} introduced \textit{EFFI-LEARNER}, a multi-iteration self-optimization framework that uses execution-time and memory-usage profiles as feedback to guide LLMs in refining their generated code. Their results demonstrated that incorporating profiling information into the code refinement process can substantially improve the efficiency of LLM-generated code. PIE~\cite{shypula2024learning} demonstrated that retrieval and adaptation techniques can enable LLMs to learn performance-improving edits to existing code, achieving substantial gains in runtime efficiency. More recently, ECO~\cite{kim2025eco} extended these approaches by providing models with both the root causes of inefficiencies and targeted optimization guidance. However, results for energy-specific improvements remain less consistent.

\section{Methodology}
{
\rowcolors{2}{gray!10}{white}
\begin{table*}[ht]
\centering
\newcolumntype{L}[1]{>{\raggedright\arraybackslash}m{#1}}
\newcolumntype{C}[1]{>{\centering\arraybackslash}m{#1}}
\renewcommand{\arraystretch}{1.2}
\begin{tabular}{
    L{3.2cm}
    L{1.0cm}
    L{1.8cm}
    L{2.1cm}
    L{1.4cm}
    C{0.8cm}
    C{0.8cm}
    C{0.8cm}
    C{2.4cm}
}
\toprule
\multicolumn{3}{c}{\textbf{Model Details}} &
\multicolumn{6}{c}{\textbf{Model Configuration}} \\
\cmidrule(lr){1-3}
\cmidrule(lr){4-9}

\textbf{Model Name} &
\textbf{Model Weights} &
\textbf{Knowledge Cut-off} &
\textbf{API Provider} &
\textbf{API Type Used} &
\textbf{Temp.} &
\textbf{Top\_p} &
\textbf{Top\_k} &
\textbf{Reasoning / Thinking Configuration} \\
\midrule
Claude 4.5 Haiku & Closed & Feb. 2025 & Anthropic & Batch &
$^{*}$1 &
$^{*}$1 &
$^{*}$0 &
$^{*}$Thinking: Disabled \\
Gemini 3.8 Flash & Closed  & Mar. 2026 & Google Vertex AI & Batch &  N/A & N/A & N/A & $^{*}$ Thinking Level: Medium \\

GPT-5.6 Terra & Closed & Feb. 2026 & OpenAI & Batch &  N/A & N/A & N/A & $^{*}$ Reasoning Effort: Medium \\

\shortstack[l]{\\GPT-OSS-120B\\(116.8B total / 5.1B active)} & Open & Jun. 2024 & Fireworks AI & Batch &  1 & 1 & $^{*}$0 & Reasoning Effort: High\\

\shortstack[l]{\\Gemma 4 26B A4B IT\\(25.2B total / 3.8B active)} & Open & Jan. 2026 & Fireworks AI & Batch &  1 & 0.95 & 64 & Thinking: Enabled\\

\shortstack[l]{\\Granite-4.0-H-Small\\(32B total / 9B active)} & Open & Apr. 2024 & 
IBM watsonx.ai & Standalone &  $^{*}$1 & $^{*}$1 & $^{*}$50 & N/A\\

\shortstack[l]{Kimi K2.7 Code \\ (FP8 Quantised) \\ (1T total / 32B active)} & Open & Not Released (Model Release: Jun. 2026) & Fireworks AI & Batch &  1 & 0.95 & $^{*}$0 & Reasoning Effort: High\\

\shortstack[l]{MiniMax M3\\(456B total / 45.9B active)} & Open &  Not Released (Model Release: Jun. 2026) & Fireworks AI & Batch &  0.95 & 0.95 & 40 & Reasoning Effort: High\\

\shortstack[l]{\\Muse Glimmer 30B\\($\sim$29.6B total)}  & Open & Jan. 2026 & Fireworks AI & Batch &  1 & 0.95 & 64 & Reasoning Effort: High \\

\shortstack[l]{ \\ Qwen3-Coder-480B-A35B\\-Instruct \\(480B total / 35B active)} & Open & Jun. 2025 & Fireworks AI & Batch &  0.7 & 0.8 & 20 & N/A \\

\bottomrule
\end{tabular}
\caption{Overview of the LLMs under study along with their respective configurations used for the study. An asterisk ($*$) indicates the default value used by the model provider, while N/A denotes a configuration that is not applicable to the model.}
\label{tab:model-comparison}
\vspace{-5mm}
\end{table*}
}
\subsection{Problem Dataset Selection}
In this study, we use the EffiBench dataset~\cite{huang2024effibench} consisting of 878 Python-solvable programming problems sourced from LeetCode~\cite{leetcode}. We chose EffiBench because it leverages LeetCode's diverse collection of algorithmic programming tasks with standardized problem statements, making it a well-established benchmark for evaluating code. Additionally, the benchmark provides a comprehensive suite of test cases for each problem, enabling rigorous and consistent assessment of the functional correctness of generated solutions. 

The benchmark spans three difficulty levels: Easy (145), Medium (510), and Hard (223). The problems also cover eleven major algorithmic categories, including Dynamic Programming (258), Greedy (231), Sorting (230), Binary Search (139), Bit Manipulation (97), Two Pointers (88), Sliding Window (62), Breadth-First Search (58), Depth-First Search (53), Backtracking (40), and Divide and Conquer (19). We also augment the dataset with an additional metadata field representing the published date of each problem. These dates are obtained from the LeetCodeDataset~\cite{xia2025leetcodedatasettemporaldatasetrobust} and span problems published between 2015 and the end of 2023. This metadata allows us to analyze model performance with respect to the temporal availability of training data.

In addition to the original Python benchmark from EffiBench, we extend the dataset to support C++ evaluation while preserving the original correctness criteria. We analyze LeetCode problem descriptions and corresponding C++ solutions on AlgoMonster to identify function signatures, invocation interfaces, and input structures. Based on this analysis, we convert Python test cases into C++-compatible formats and define standardized representations for complex data structures, such as linked lists, graphs, and trees. We also include the required headers and boilerplate to compile and execute LLM-generated solutions, enabling consistent evaluation and fair cross-language comparisons.

\begin{figure}[t]
    \centering
    \fbox{\includegraphics[width=0.95\linewidth,
        height=40mm]
        {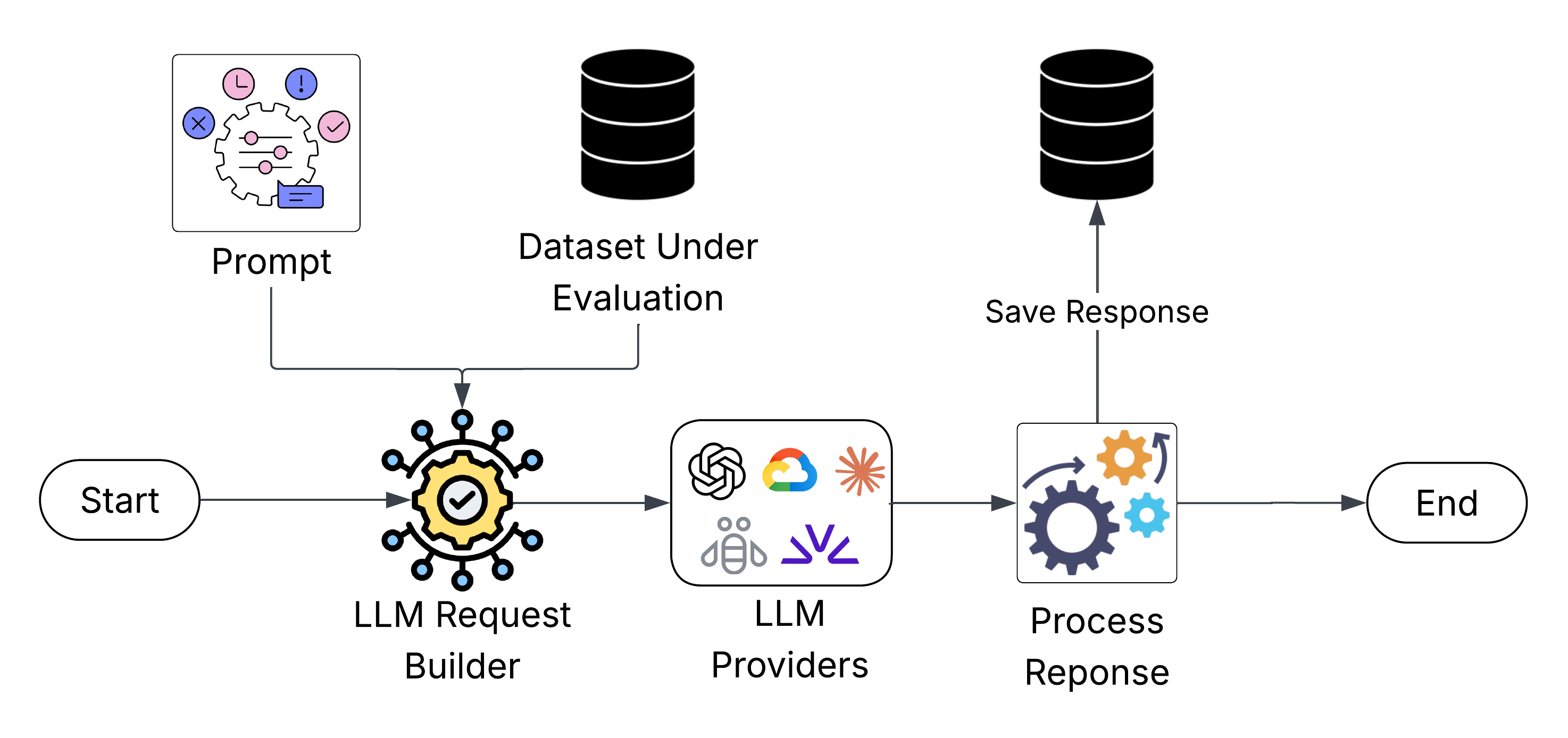}}
    \caption{Code generation workflow applied to each prompting strategy across both evaluation stages.}
    \label{fig:code_generation_flow}
    \vspace{-3mm}
\end{figure}

\subsection{LLMs Under Study}
\subsubsection{Selected LLMs}
This study evaluates three proprietary large language models from the widely used model providers—\mbox{GPT-5.6 Terra}~\cite{openaigpt5.6terra} from OpenAI, \mbox{Gemini 3.8 Flash}~\cite{gemini3.8flashmodelcard} from Google DeepMind, and \mbox{Claude 4.5 Haiku}~\cite{claude4.5modelcard} from Anthropic  —as well as seven popular open-weight models: \mbox{Granite-4.0-H-Small}~\cite{ibm2025granite40hsmall}, \mbox{Gemma 4 26B A4B IT}~\cite{gemmateam2026gemma4}, \mbox{Muse Glimmer 30B}~\cite{meta2026museglimmer}, GPT-OSS-120B~\cite{openai2025gptossmodelcard}, \mbox{Kimi K2.7 Code}~\cite{moonshotai2026kimik27code}, \mbox{MiniMax M3}~\cite{minimax2026m3modelcard}, and Qwen3-Coder-480B-A35B-Instruct~\cite{qwen2025qwen3coder480ba35b}. To provide context for the evaluation and facilitate reproducibility, we document knowledge cutoff dates and the model configurations including API provider and model-specific inference-time generation parameters for each evaluated model. 
The knowledge cutoff dates help assess the likelihood that the benchmark problems and their canonical solutions were included in the models' training corpora, providing important context for interpreting the results. Since our benchmark includes problems published through the end of 2023, it is likely that these models were exposed to most or all of the benchmark problems and their canonical solutions during training. 
\subsubsection{LLM Configuration}
The inference parameters were configured according to the recommended settings provided by each model provider, where available. We also leveraged batch inference APIs and the sampling parameter $n$, where supported, to reduce the number of API calls and improve inference efficiency.

For \mbox{GPT-5.6 Terra}, \mbox{Claude 4.5 Haiku}, \mbox{Gemini 3.8 Flash}, and Granite-4.0-H-Small, we used the default configuration settings provided by their respective model providers. \mbox{GPT-5.6 Terra} was accessed through the \texttt{/v1/batches} API, with each batch containing requests to the \texttt{/v1/chat/completions} endpoint. For \mbox{Claude 4.5 Haiku}, we used the \texttt{/v1/messages/batches} API for all inference requests, while \mbox{Gemini 3.8 Flash} was accessed through the \texttt{/batchPredictionJobs} API. For \mbox{Granite-4.0-H-Small}, inference was performed through IBM watsonx.ai using the \texttt{ml/v1/text/chat} endpoint, as batch inference was not supported by the platform.

For \mbox{Gemma 4 26B A4B IT}, \mbox{Muse Glimmer 30B}, GPT-OSS-120B, Kimi K2.7 Code, MiniMax M3, and Qwen3-Coder-480B-A35B-Instruct, which were accessed through Fireworks AI, we used the \texttt{/batchInferenceJobs} API to perform inference and obtain the required samples. For Qwen3-Coder-480B-A35B-Instruct, we set the \texttt{repetition\_penalty} to $1.05$, following the recommendation provided by the model provider.

The complete details of the evaluated models, including their providers, access mechanisms, and inference configurations, are summarized in Table~\ref{tab:model-comparison}.

\begin{figure}[t]
    \centering
\fbox{\includegraphics[width=0.95\linewidth,
        height=40mm]
        {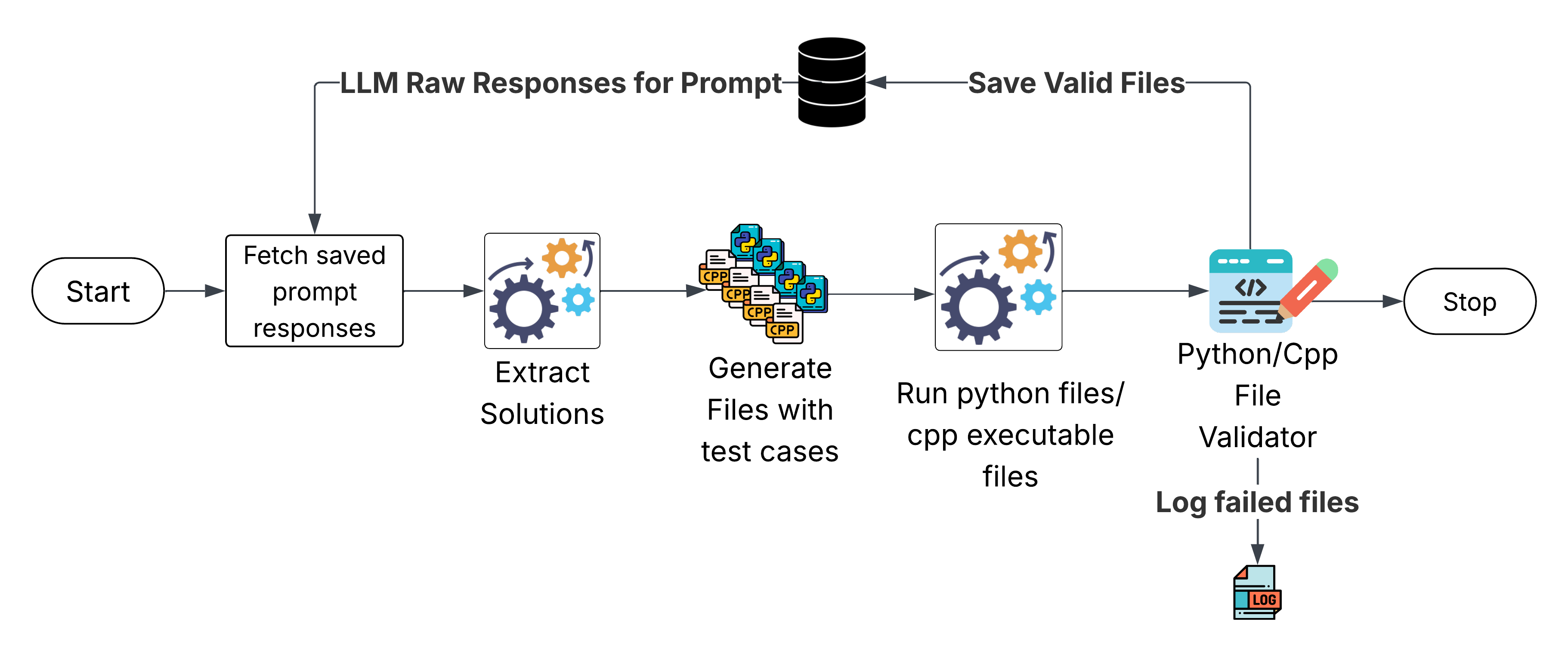}}
    \caption{Code evaluation workflow for the generated Python and C++ programs.}
    \label{fig:python_cpp_code_evaluation_flow}
    \vspace{-3mm}
\end{figure}

\subsection{LLM Code Generation}
\begin{figure}[b]
    \centering    \includegraphics[width=\linewidth,
        height=\textheight,
        keepaspectratio]
        {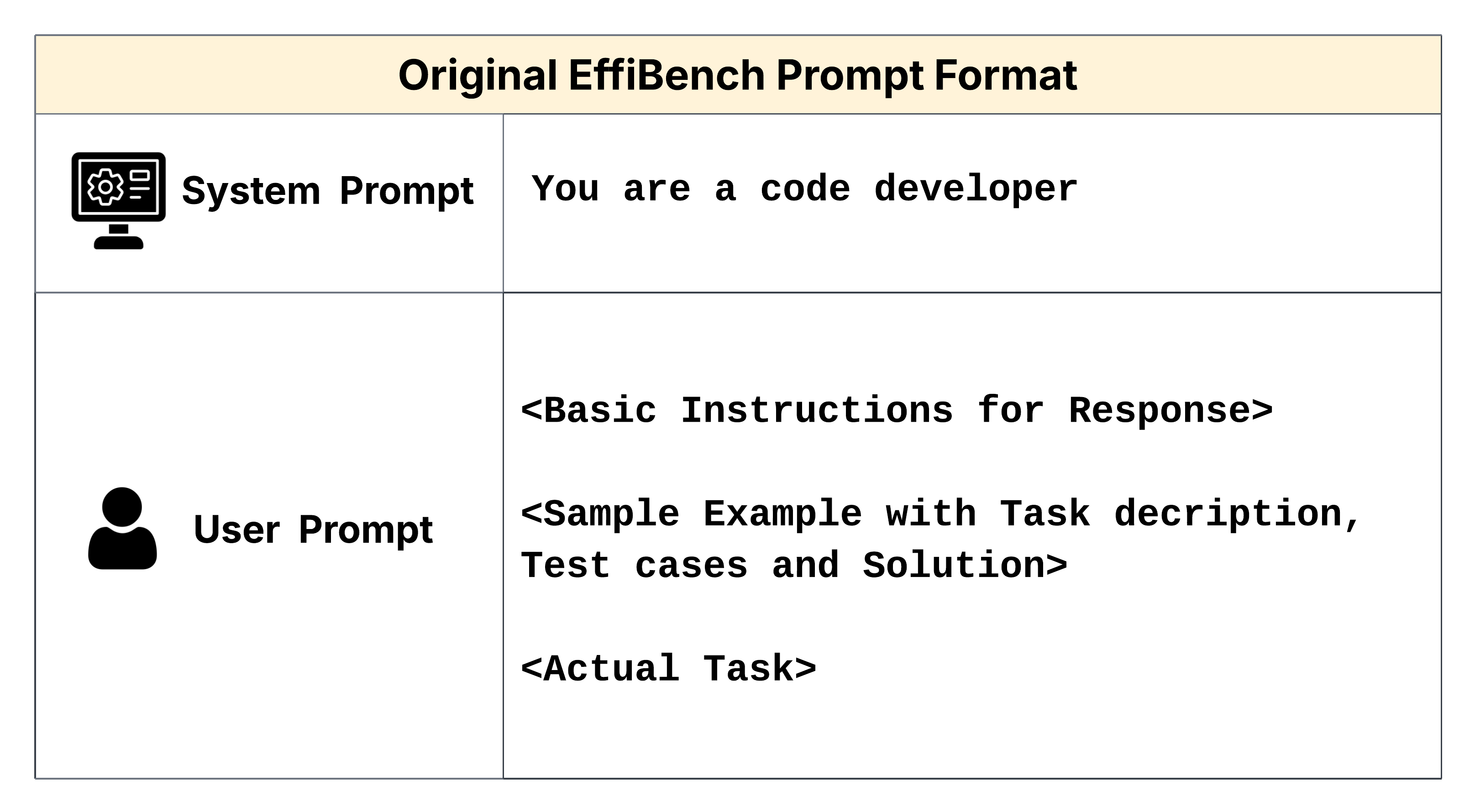}
    \caption{EffiBench prompt structure used as the baseline.}
    \label{fig:original_effibench_prompt}
    \vspace{-3mm}
\end{figure}

For each LLM, we constructed API requests in accordance with the specifications of its respective provider and API interface. Proprietary LLMs were accessed through their respective providers using the OpenAI, Anthropic, and Google Vertex AI platforms, while open-weight LLMs were accessed through Fireworks AI and IBM watsonx.ai. Our evaluation was conducted in two stages. In the initial analysis, we evaluated 21 prompting strategies across the selected models on a subset of 110 problems, using the EffiBench baseline prompt (Figure~\ref{fig:original_effibench_prompt})as the reference. This stage was designed to identify prompting strategies that consistently improved the energy efficiency of generated code. Based on the results of this analysis, 8 prompting strategies were subsequently evaluated on the extended EffiBench dataset, along with the baseline prompt, to assess their effectiveness across a broader set of programming problems.

To account for the stochastic nature of LLM generation, each problem was evaluated using 10 independent code samples per model and prompt. For multi-turn prompting strategies, these 10 samples were generated only for the base prompt, and subsequent prompting turns were applied to the resulting samples rather than generating additional independent samples. To ensure consistency across multi-turn evaluations, only the response generated in the preceding turn was passed to the subsequent turn, without including the model's intermediate reasoning context, where applicable.

All LLM-generated responses from both evaluation stages were retained for downstream analysis, including functional correctness verification and fine-grained energy efficiency benchmarking. The high-level code generation workflow is illustrated in Figure~\ref{fig:code_generation_flow}.

\begin{figure*}[t]
    \centering
    \includegraphics[width=\textwidth]{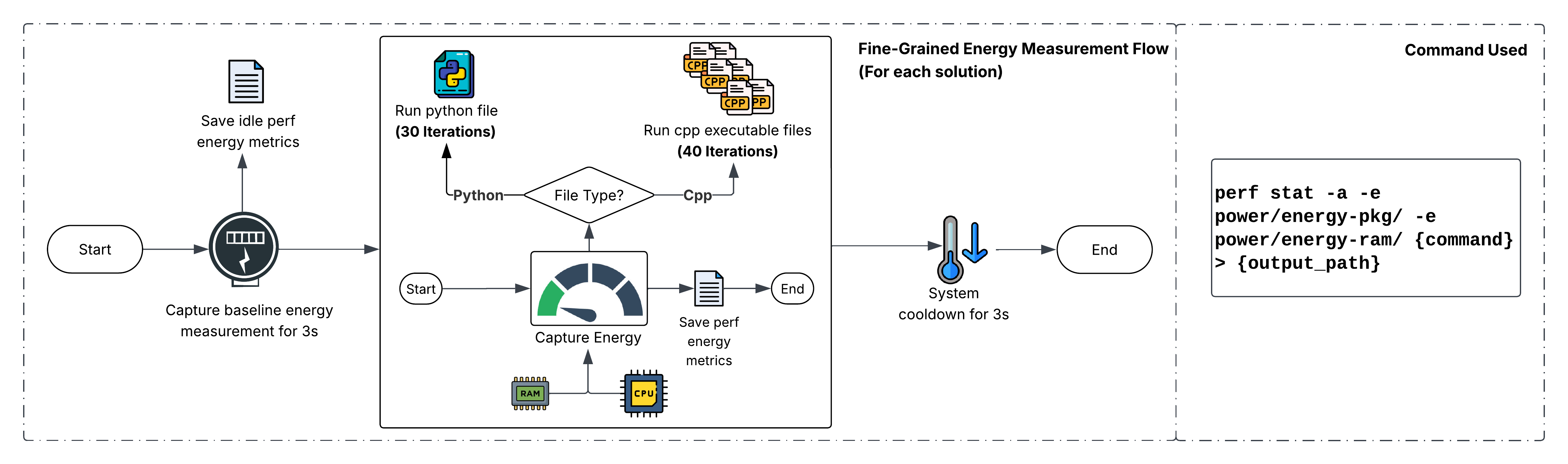}
    \caption{Fine-grained energy measurement process for Python and C++ programs.}
    \label{fig:python_cpp_energy_measurement_flow}
\end{figure*}

\subsection{LLM Code Evaluation}
Following code generation using the predefined experimental setup, the generated outputs undergo a code extraction and evaluation pipeline. The language-specific markers included in the prompts facilitate reliable extraction of the generated code snippets for each sample across all prompting strategies.

For Python, the extracted code is combined with the required import statements and assertion-based test cases to construct separate Python scripts for the small and large test suites. Each script is executed with a timeout of 120 seconds. A generated solution is considered correct only if it successfully passes all the small test cases and 100 large test cases within the specified time limit. The Python scripts corresponding to the correct solutions are subsequently retained for the energy measurement phase.

For C++, the extracted code is first augmented with the necessary header files. Subsequently, a separate C++ source file is generated for each test case associated with the problem, resulting in 100 source files per generated solution. Each source file is compiled into its corresponding executable, which is then executed with a timeout of 10 seconds. Similar to the Python evaluation, a solution is classified as correct only if all executables successfully pass their respective test cases within the allotted time. The executables corresponding to the large test cases for these correct solutions are then preserved for subsequent energy measurements. Figure~\ref{fig:python_cpp_code_evaluation_flow} shows a high-level code evaluation workflow.

\subsection{Energy Measurement Setup}
To ensure a fair comparison between the proposed prompting strategies and the baseline prompt for each LLM, all validated solutions were evaluated on the same hardware instance under identical execution conditions.

Energy consumption for each generated solution was measured using the \texttt{perf stat} command~\cite{perfstat}. Since \texttt{perf stat} reports package (PKG) and DRAM energy consumption at the system-wide level, an idle baseline was first recorded by leaving the system idle for 3 seconds. The corresponding language-specific solution was then executed while collecting energy measurements. Each validated Python solution was executed 30 consecutive times, whereas all validated C++ executables associated with a solution were executed 40 consecutive times. These repetition counts were selected to ensure that the total measurement duration exceeded 1 second, thereby improving the reliability of hardware energy counter sampling and reducing inaccuracies associated with profiling very short-running programs. A 3-second cooldown interval was also introduced between consecutive measurements of each solution to allow the system to return to a stable thermal and power state.

{
\rowcolors{2}{gray!8}{white}
\begin{table*}[p]
\centering
\newcolumntype{L}[1]{>{\raggedright\arraybackslash}m{#1}}
\newcolumntype{C}[1]{>{\centering\arraybackslash}m{#1}}
\renewcommand{\arraystretch}{1.2}
\begin{tabular}{
    C{2.5cm}
    C{1.7cm}
    C{1.2cm}
    L{2.5cm}
    L{2.2cm}
    L{5.5cm}
}
\toprule
\textbf{Prompt Number} & \textbf{Conversation Type} &  \textbf{Source} & \textbf{Prompt Type} & \textbf{Change Scope} & \textbf{Additions Made to EffiBench Prompt} \\
\midrule
\rowcolor{gray!30}
\textbf{$\star$Prompt 0 (EffiBench Prompt)}  & \textbf{Single\-Turn} & \textbf{~\cite{huang2024effibench}} & \textbf{Zero-Shot} & \textbf{-} & \textbf{-} \\

Prompt 1  & Single\-Turn & ~\cite{cappendijk2024generating} ~\cite{apsan2026generating} ~\cite{tuttle2024generating} ~\cite{cursaru2024controlled}& Zero-Shot & User Prompt & Energy-Efficiency Keyword \\

Prompt 2  & Single-Turn & - & Zero-Shot & User Prompt &  Energy-Efficiency Keyword and Definition \\

Prompt 3  & Single-Turn & - & Few-Shot & User Prompt & Energy-Efficiency Keyword, General Efficiency Guidelines \\

\rowcolor{green!10}
\textbf{$\star$ Prompt 4}  & Single-Turn & - & Persona-Based + Zero-Shot & User Prompt & General Efficiency Guidelines, Language-Specific Efficient Practices \\

Prompt 5  & Single-Turn & - & Persona-Based + Zero-Shot & System Prompt &  Expert Competitive Programmer Persona, Energy-Efficiency Definition  \\

Prompt 6  & Single-Turn & - & Chain-of-Thought & User Prompt & Explicitly prompting the model to reason,  Added efficiency explanation to the sample code \\

Prompt 7  & Single-Turn & - & Chain-of-Thought & User Prompt & Energy-Efficiency Keyword, General Efficiency Guidelines, 1 Energy-Efficient Example with Explanation \\

Prompt 8  & Single-Turn & - & Few-Shot & User Prompt & Prompting for optimal solution explicitly, 1 example with energy-efficient and inefficient Solutions and explanation \\

Prompt 9  & Single-Turn & - & Persona-Based + Few-Shot & System Prompt + User Prompt & Expert Competitive Programmer Persona, Energy-Efficiency Definition,  2 Energy-Efficient Solutions with Task Descriptions \\

\rowcolor{green!10} \textbf{$\star$ Prompt 10} & Single-Turn & - & Persona-Based + Few-Shot & System Prompt + User Prompt & Expert Competitive Programmer Persona, Energy-Efficiency Definition,  3 LeetCode solutions(efficient \& inefficient),  with task descriptions external to dataset under study \\

\rowcolor{green!10} \textbf{$\star$ Prompt 11} & Single-Turn & - & Persona-Based + Few-Shot & System Prompt + User Prompt & Expert Competitive Programmer Persona, Energy-Efficiency Definition, General Efficiency Guidelines, 11 LeetCode solutions (efficient \& inefficient) with task descriptions added from each algorithm category, external to dataset under study \\

\rowcolor{green!10} \textbf{$\star$ Prompt 12} & Single-Turn & ~\cite{ashraf2025greencode} & Persona-Based & System Prompt & Senior Software Engineer with 10 years of Experience Persona \\

\rowcolor{green!10} \textbf{$\star$ Prompt 13 (Prompt 0 - EE)} & \shortstack[c]{Multi-Turn\\(2 Turns)} & ~\cite{niu2024efficiency} & Chain-of-Thought & User Prompt & Prompt 0  at Turn 1, Prompting for Energy-Efficiency at Turn 2 \\

Prompt 14 & \shortstack[c]{Multi-Turn\\(3 Turns)} & ~\cite{niu2024efficiency} & Chain-of-Thought & User Prompt & Prompt 0 at Turn 1, Prompting for generating energy-efficiency strategy at Turn 2, Prompting for applying strategy at Turn 3 \\

Prompt 15 & \shortstack[c]{Multi-Turn\\(3 Turns)} & ~\cite{peng2025perfcodegen} & Chain-of-Thought & User Prompt & Prompt 0 at Turn 1, Prompting for analyzing Big(O) time at Turn 2, Prompting for optimization based on analysis at Turn 3 \\

Prompt 16 & Single-Turn & ~\cite{jonnala2025measuring} & Chain-of-Thought & User Prompt & Explicitly requests chain-of-thought reasoning and adds reasoning for the sample code provided in Prompt 0. \\

Prompt 17 & Single-Turn & - & Persona-Based + Few-Shot & System Prompt + User Prompt & Prompt 11 with Senior Software Engineer with 10 years of Experience Persona \\

\rowcolor{green!10} \textbf{$\star$ Prompt 18} & Single-Turn & - & Persona-Based + Few-Shot & System Prompt + User Prompt & Prompt 10 with Senior Software Engineer with 10 years of Experience Persona\\

Prompt 19 & Single-Turn & - & Persona-Based + Few-Shot & User Prompt & Senior Software Engineer with 10 years of Experience Persona, Prompt 4 Changes \\

\rowcolor{green!10} \textbf{$\star$ Prompt 20 (Prompt 4 - EE)} & \shortstack[c]{Multi-Turn\\(2 Turns)} & - & Persona-Based + Zero-Shot + Chain-of-Thought & User Prompt & Prompt 4 at Turn 1, Prompting for Energy-Efficiency at Turn 2 \\

\rowcolor{green!10} \textbf{$\star$ Prompt 21 (Prompt 10 - EE)} & \shortstack[c]{Multi-Turn\\(2 Turns)} & - & Persona-Based + Few-Shot + Chain-of-Thought & System Prompt + User Prompt & Prompt 10 at Turn 1, Prompting for Energy-Efficiency at Turn 2 \\

\bottomrule
\end{tabular}
\caption{Summary of 21 prompting strategies and the EffiBench prompt evaluated in the initial analysis. Strategies marked with \textbf{$\star$} and highlighted in green were selected for the final evaluation for Python and C++ code generation across our extended EffiBench dataset and all LLMs under evaluation.}
\label{tab:20-prompt-summary}
\end{table*}
}

The net energy consumption of each solution was calculated by subtracting the estimated idle energy from the total measured energy. Specifically, the idle energy per second was multiplied by the total execution time across all iterations to estimate the baseline energy consumption during execution. This baseline was then subtracted from the measured energy, and the resulting net energy was divided by the number of iterations to obtain the average energy consumption per execution. To reduce measurement noise and improve result reliability, the entire measurement procedure was repeated twice, and the final reported energy consumption was calculated as the average across the two repetitions. 

Figure~\ref{fig:python_cpp_energy_measurement_flow} presents the fine-grained framework used to measure the energy consumption of the LLM-generated Python and C++ solutions for each prompting strategy.

\begin{figure*}[!t]
    \centering

    \begin{subfigure}{\linewidth}
        \centering
        \includegraphics[
            width=\linewidth,
            height=0.90\textheight,
            keepaspectratio
        ]{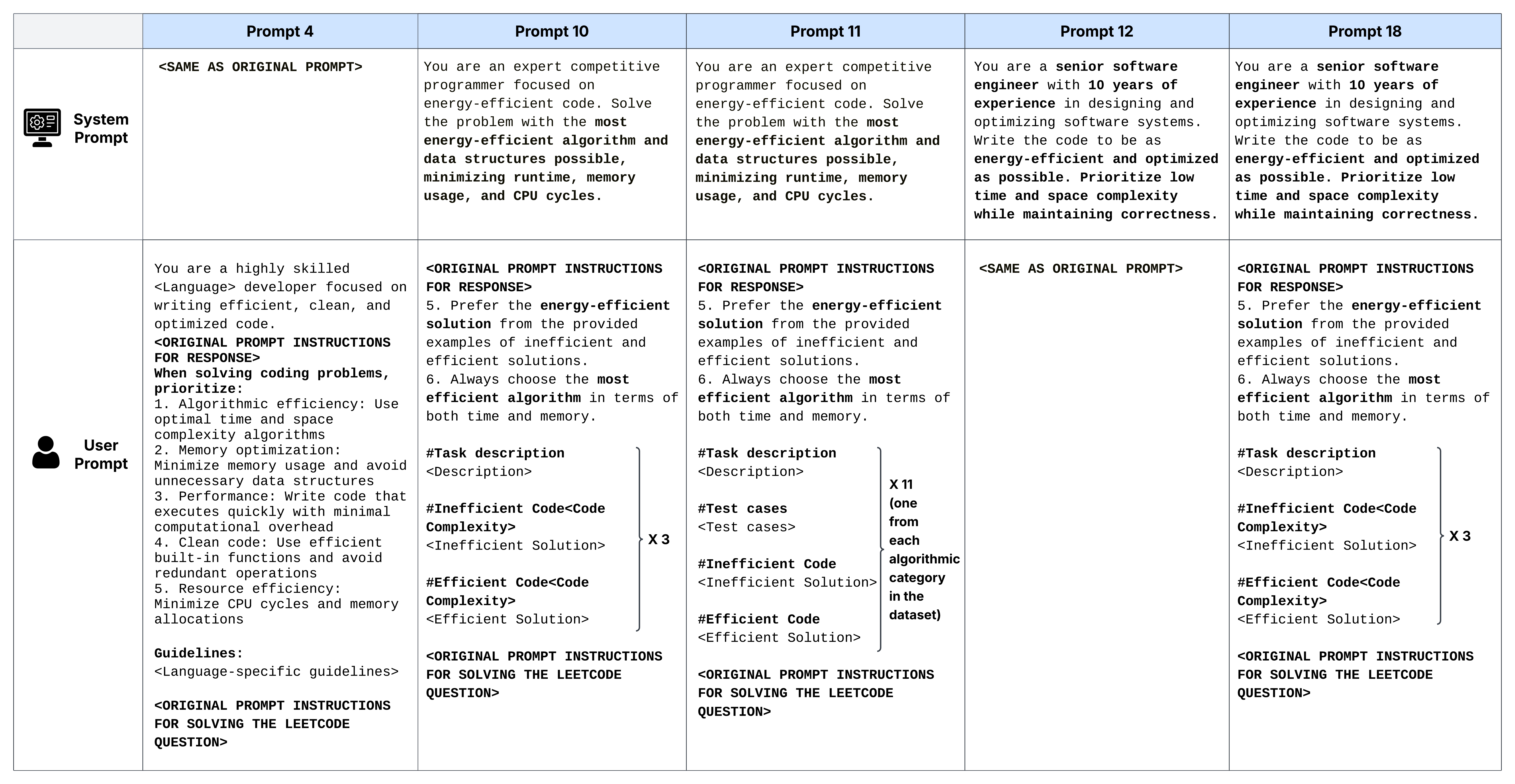}
        \caption{Five selected single-turn prompt strategies.}
        \label{fig:single_turn_finalised_ee_prompt}
    \end{subfigure}

    \vspace{3mm}

    \begin{subfigure}{\linewidth}
        \centering
        \includegraphics[
            width=\linewidth,
            height=0.40\textheight,
            keepaspectratio
        ]{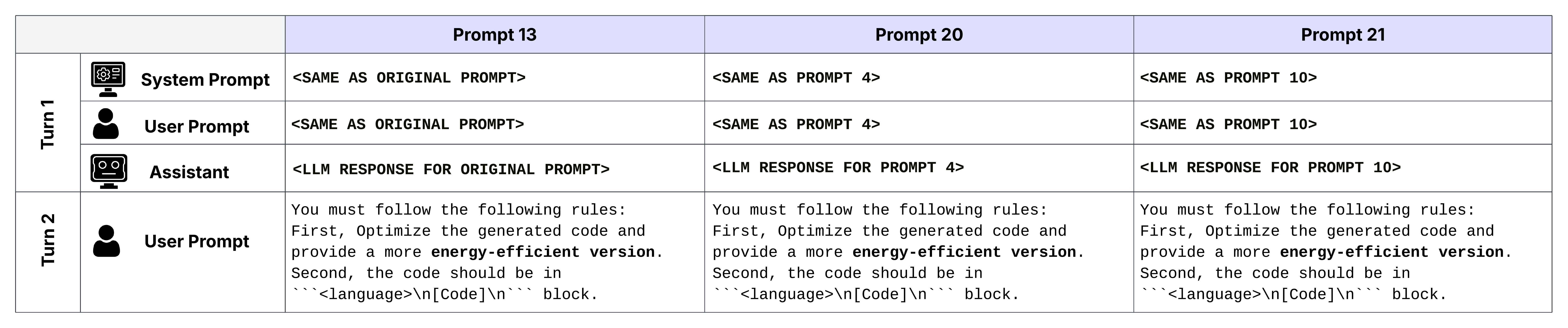}
        \caption{Three selected multi-turn prompt strategies.}
        \label{fig:multi_turn_finalised_ee_prompt}
    \end{subfigure}

    \caption{Finalized prompting strategies for Python and C++ code generation. 
    The top panel presents the five single-turn prompting strategies, while the 
    bottom panel presents the three multi-turn prompting strategies. Both are 
    applied across the extended EffiBench dataset and evaluated across all LLMs 
    under study.}
    \label{fig:finalised_ee_prompts}
\end{figure*}

\subsection{Environment Setup}
Energy measurements were conducted using Intel Skylake and Cascade Lake instances on Chameleon Cloud~\cite{chameleon_cloud}. These instances provide Model-Specific Registers (MSR) that enable the perf~\cite{perfstat} tool to read both package (pkg) and RAM energy values. The operating system used was Ubuntu 24.04, the Python interpreter version was 3.12.3, and the C++ compiler version was 13.3.0. Additionally, the C++ solutions were compiled using the GNU C++ compiler with the C++20 standard and the \texttt{-O2} optimization flag. All measurements for solutions to common problems across prompting strategies within each model were performed on the same hardware configuration to ensure consistency and comparability.

\subsection{Evaluating Prompting Strategies for Energy-Efficient Code Generation}
To evaluate prompting strategies that can improve upon the baseline prompt, we used the EffiBench prompt as the baseline. The EffiBench prompt was selected because it provides minimal instructions while clearly specifying the expected output format, thereby providing a simple and consistent reference point for evaluating the effect of additional prompting strategies.

To identify prompting strategies that consistently encourage the generation of energy-efficient solutions, we developed prompts based on intuitive approaches to guide LLMs toward more energy-efficient code generation. We also reviewed prior research on prompt engineering and incorporated relevant techniques into our prompt designs. We restricted our selection to strategies that rely solely on the capabilities of the LLM and do not require external tools, systems, or feedback mechanisms to guide code generation. Based on these criteria, we developed 21 prompting strategies for evaluation, incorporating techniques adapted from existing research along with novel strategies proposed in this work. These strategies are summarized in Table~\ref{tab:20-prompt-summary}.

\subsection{Finalised Prompting Techniques for Energy-Efficient Code Generation}

Based on the results of the initial analysis, we selected 8 energy-efficient prompting strategies (Prompts 4, 10, 11, 12, 13, 18, 20, and 21) for final evaluation against the baseline EffiBench prompt, highlighted in Table~\ref{tab:20-prompt-summary}. Of these 8 variants, five were single-turn strategies (Prompts 4, 10, 11, 12, and 18), while three were multi-turn strategies (Prompts 13, 20, and 21). The multi-turn strategies extend the baseline or selected single-turn prompts (Prompts 0, 4, and 10, respectively) with an additional energy efficiency refinement step. Separate prompt instances are constructed for Python and C++, with the only difference being the language-specific reference implementation.

The baseline prompt, adopted from EffiBench~\cite{huang2024effibench}, consists of a system prompt and a user prompt. The system prompt assigns the LLM the role of a code developer, while the user prompt specifies the required output format, instructs the model to generate code that passes the provided sample test cases, and includes a worked example comprising a task description, sample test cases, and a reference language-specific solution. This example is followed by the actual programming task to be solved. The complete prompt format is illustrated in Figure~\ref{fig:original_effibench_prompt}.

Among the five single-turn strategies, Prompt 4 extends the baseline user prompt by adopting the persona of a skilled Python/C++ developer and incorporating both general code-efficiency guidelines and language-specific best practices. The general guidelines instruct the LLM to prioritize algorithmic and memory efficiency, as well as performance and resource efficiency, while the language-specific guidelines recommend libraries and programming patterns intended to improve efficiency.
Prompt 10 modifies both the system and user prompts. The system prompt assigns the LLM the role of an expert competitive programmer and provides a definition of energy efficiency. The user prompt presents three LeetCode-style problems with their corresponding task descriptions, along with examples of energy-efficient and inefficient code and their associated computational complexities.
Prompt 11 retains the system prompt used in Prompt 10 but expands the user prompt to include 11 LeetCode-style problem descriptions, with one problem representing each algorithmic category in the dataset. Each example includes both an efficient and an inefficient solution.
The example problems used in Prompts 10 and 11 are drawn from outside the evaluation dataset and therefore do not overlap with the problems used for evaluation.
Prompt 12 is adapted from Ashraf et al.~\cite{ashraf2025greencode} and modifies the system prompt by assigning the LLM the role of a Senior Software Engineer with 10 years of experience.
Finally, Prompt 18 modifies both the system and user prompts. Similar to Prompt 12, its system prompt assigns the LLM the role of a Senior Software Engineer with 10 years of experience. Its user prompt, similar to Prompt 10, provides three LeetCode-style problems with their task descriptions and examples of efficient and inefficient code, in addition to the problem to be solved. The five single-turn prompting strategies described above are illustrated in Figure~\ref{fig:single_turn_finalised_ee_prompt}.

The multi-turn strategies, Prompts 13, 20, and 21, are inspired by the approach proposed in Niu et al~\cite{niu2024efficiency}. These strategies extend the responses generated by the corresponding base prompts, Prompts 0, 4, and 10, respectively, by introducing a second turn that explicitly instructs the LLM to generate \textit{energy-efficient code}. The three single-turn prompting strategies described above are illustrated in Figure~\ref{fig:multi_turn_finalised_ee_prompt}.

The prompting strategies used in this paper and the experimental results are open-sourced at \url{https://github.com/didclab/effective-prompting-for-energy-efficient-code/}.
\section{Evaluation and Analysis}

\subsection{Initial Analysis for Prompt Selection}
\begin{figure}[t]
    \centering
    \fbox{\includegraphics[width=0.95\linewidth,
        height=0.9\textheight,
        keepaspectratio]
        {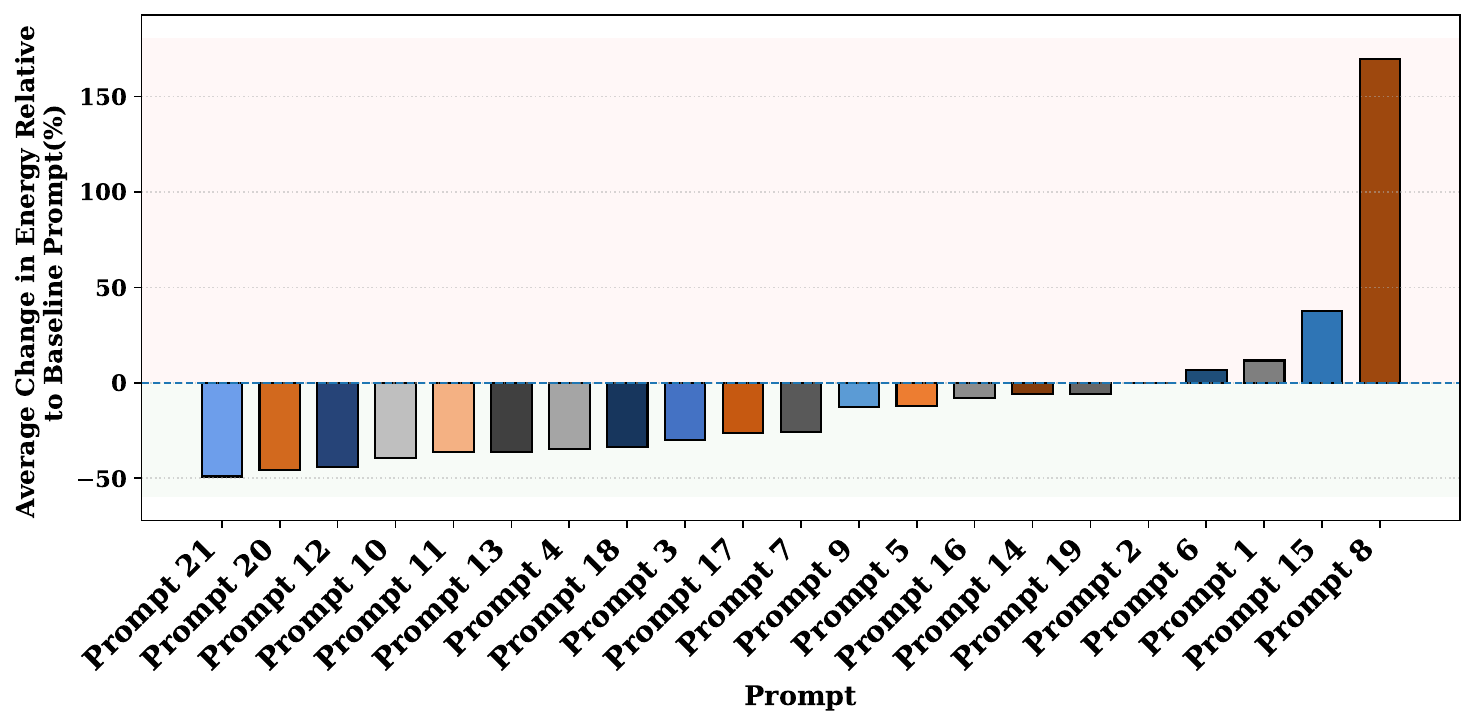}}
    \caption{Average change in energy consumption relative to the baseline prompt across 110 python problems and 21 prompting strategies, aggregated across the 3 LLMs evaluated in the initial analysis.}
    \label{fig:initial_analysis_prompts_impact}
\end{figure}
Due to the computational cost of evaluating all 878 benchmark problems across multiple prompting strategies, we constructed a stratified subset of 110 problems, selecting 10 problems from each of the 11 algorithmic categories to ensure representative coverage of the problem space. Using the EffiBench baseline prompt (Figure~\ref{fig:original_effibench_prompt}) as the reference, we evaluated 21 prompting strategies across three models: \mbox{Kimi K2.7 Code}, \mbox{MiniMax M3}, and \mbox{Granite-4.0-H-Small}. This initial screening was conducted exclusively on Python problems, under the assumption that the effectiveness of prompting strategies would remain consistent across programming languages.

Following code generation, functional correctness evaluation, and energy benchmarking, we applied filtering criteria to establish a consistent set of problems for prompt-selection analysis. First, we retained only those problems for which all three models produced correct solutions across the evaluated prompting strategies. Additionally, given the relatively small size of the initial evaluation set, we required each problem to have at least two correct solutions generated by the model. Based on these criteria, we identified 59 problems for which all 21 prompting strategies and the baseline prompt produced functionally correct solutions across all three models. This filtered set was subsequently used to identify prompting strategies that consistently reduced the energy consumption of generated code. 

As shown in Figure~\ref{fig:initial_analysis_prompts_impact}, Prompts 21 ($-49\%$), 20 ($-45\%$), 12 ($-44\%$), 10 ($-39\%$), 11 ($-36\%$), 13 ($-36\%$), 4 ($-35\%$), and 18 ($-34\%$) achieved the largest reductions in energy consumption relative to the baseline prompt. These 8 prompting strategies were therefore selected for evaluation on the extended EffiBench dataset across all ten models and for both Python and C++ code generation.

\begin{table*}[t]
\label{tab:python-difficulty-algorithm}
\resizebox{\textwidth}{!}{%
\begin{tabular}{lcccccccccccc}
\toprule
& \multicolumn{12}{c}{\textbf{Algorithms}} \\
\cmidrule(lr){2-13}
\textbf{Difficulty} &
\textbf{Backtracking} &
\textbf{BFS} &
\textbf{Binary Search} &
\textbf{Bit Manip.} &
\textbf{DFS} &
\textbf{Divide \& Conquer} &
\textbf{DP} &
\textbf{Greedy} &
\textbf{Sliding Window} &
\textbf{Sorting} &
\textbf{Two Pointers} &
\textbf{Total} \\
\midrule
\textbf{Easy} & 1 & 0 & 13 & 21 & 0 & 2 & 7 & 25 & 7 & 47 & 24 & \cellcolor{gray!30}110 \\
\textbf{Medium} & 17 & 21 & 48 & 33 & 24 & 3 & 96 & 98 & 31 & 83 & 33 & \cellcolor{gray!30}333 \\
\textbf{Hard} & 3 & 6 & 24 & 7 & 2 & 4 & 45 & 18 & 8 & 17 & 0 & \cellcolor{gray!30}94 \\
\textbf{Total} & \cellcolor{gray!10}\textbf{21} & \cellcolor{gray!10}\textbf{27} & \cellcolor{gray!10}\textbf{85} & \cellcolor{gray!10}\textbf{61} & \cellcolor{gray!10}\textbf{26} & \cellcolor{gray!10}\textbf{9} & \cellcolor{gray!10}\textbf{148} & \cellcolor{gray!10}\textbf{141} & \cellcolor{gray!10}\textbf{46} & \cellcolor{gray!10}\textbf{147} & \cellcolor{gray!10}\textbf{57} & \cellcolor{gray!30}\textbf{537} \\

\bottomrule
\end{tabular}
}
\caption{Distribution of benchmark problems solved by all 10 evaluated LLMs across difficulty levels and algorithm categories for Python code generation.}
\label{tab:python_dist}
\vspace{-3mm}
\end{table*}

\begin{figure*}[!t]
    \centering
    \fbox{\includegraphics[width=0.98\textwidth]{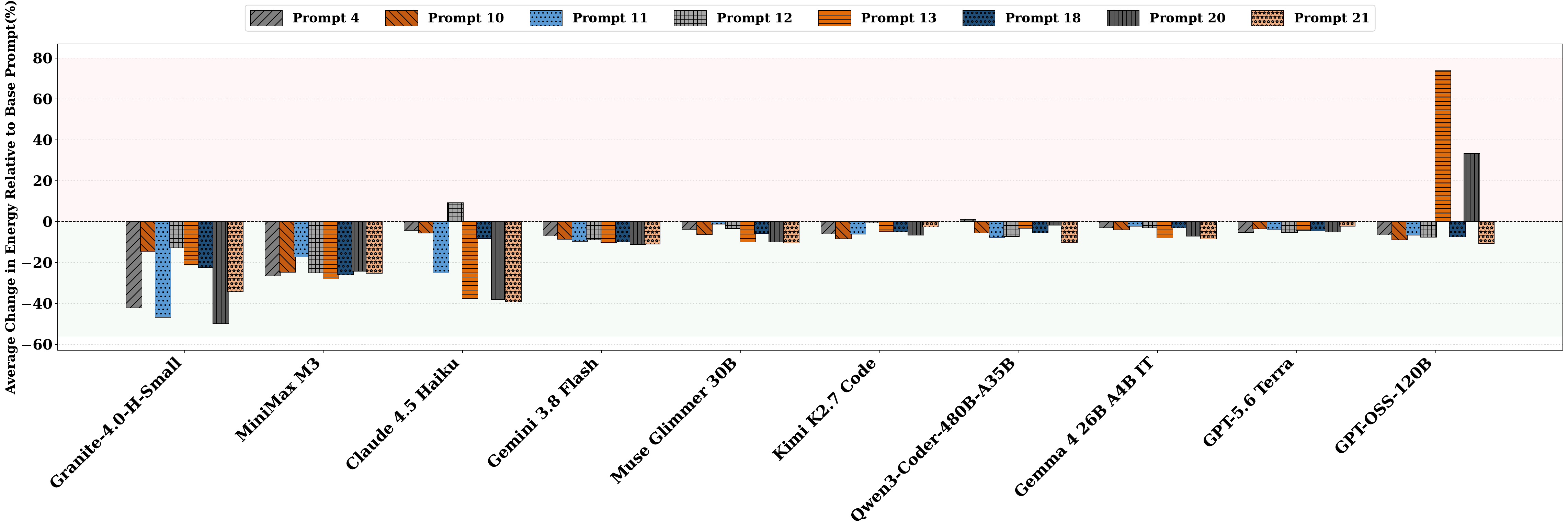}}
    \caption{Average percentage energy reduction achieved by the 8 prompting strategies relative to the baseline prompt for Python code generation. Negative values indicate lower energy consumption than the baseline prompt.}
    \label{fig:python_energy_reductions}
\end{figure*}
\subsection{Analysis of 8 Energy-Efficient Prompting Strategies Across all Models}
The 8 selected prompting strategies, together with the baseline prompt, were subsequently evaluated on our extended EffiBench dataset across all LLMs under study. To ensure a fair and consistent comparison, we restricted the analysis to problems for which all evaluated models produced functionally correct solutions under every prompting strategy for each programming language. This resulted in a common benchmark comprising 537 Python problems and 607 C++ problems. In this shared evaluation set, the proposed prompting strategies reduced energy consumption for several models compared to the baseline prompt. For Python code generation, all evaluated models showed improvements, with \mbox{Granite-4.0-H-Small}, \mbox{Claude 4.5 Haiku}, and \mbox{MiniMax M3} achieving the largest reductions (Fig.~\ref{fig:python_energy_reductions}). 
For C++ code generation, Granite-4.0-H-Small, Qwen3-Coder-480B-A35B-Instruct, \mbox{GPT-5.6 Terra}, \mbox{Muse Glimmer 30B}, and \mbox{GPT-OSS-120B} showed higher overall energy reductions, as shown in Fig.~\ref{fig:cpp_energy_reductions}.
However, the effects of prompting varied across models, programming languages, and prompting strategies. For a given model and programming language, different prompting strategies could have opposite effects, with some reducing energy consumption and others increasing it relative to the baseline. For example, as shown in Fig.~\ref{fig:python_energy_reductions}, for Python code generation, \mbox{GPT-OSS-120B} achieved an energy reduction with Prompts 4, 10, 11, 12, 18, and 21, while Prompts 13 and 20 resulted in higher energy consumption than the baseline. In addition, the same model could exhibit different energy-consumption trends across Python and C++ code generation. For example, \mbox{Gemma 4 26B A4B IT} showed reduced energy consumption for Python code generation but increased energy consumption for C++ code generation.

The following sections provide a detailed analysis of the impact of each prompting strategy across models and programming languages:

\subsection{Evaluation of Prompts on Python Code Generation}

\subsubsection{Dataset Distribution}
The common Python evaluation set comprises 537 problems that were solved correctly by all ten evaluated models across every prompting strategy. Of these, 333 problems are of Medium difficulty, while 110 are Easy and 94 are Hard. The benchmark spans a diverse range of algorithmic categories, with the largest representation coming from Dynamic Programming (148), Sorting (147), and Greedy (141). Other well-represented categories include Binary Search (85), Bit Manipulation (61), and Two Pointers (57).
The detailed data distribution can be found in Table~\ref{tab:python_dist}.

\begin{figure*}[!t]
    \centering

    \begin{subfigure}[t]{0.48\linewidth}
        \centering
        \fbox{%
            \includegraphics[
                width=0.96\linewidth,
                height=0.50\textheight,
                keepaspectratio
            ]{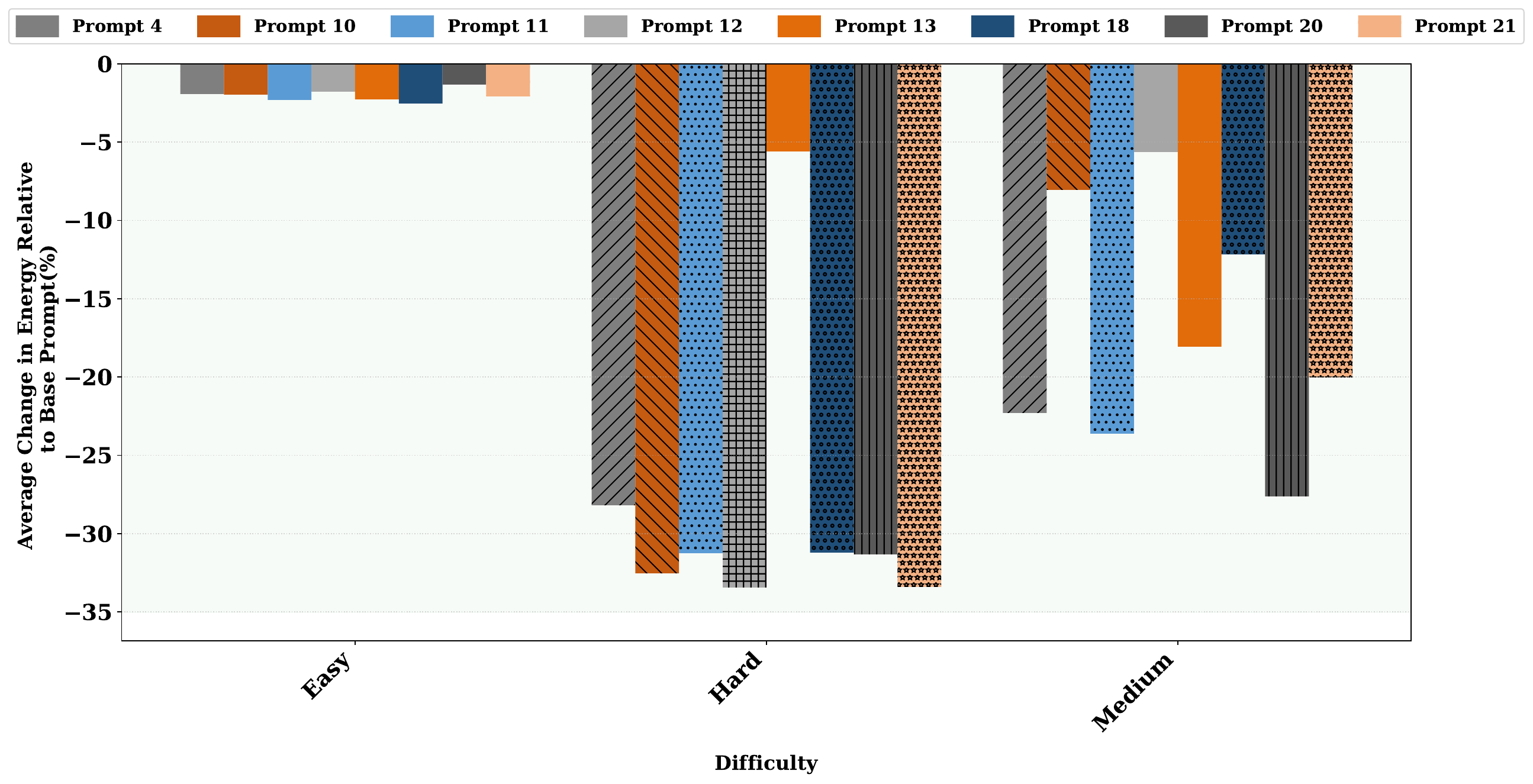}
        }
        \caption{Across problem difficulty levels}
        \label{fig:python_difficulty_prompt_impact}
    \end{subfigure}
    \hfill
    \begin{subfigure}[t]{0.48\linewidth}
        \centering
        \fbox{%
            \includegraphics[
                width=0.96\linewidth,
                height=0.50\textheight,
                keepaspectratio
            ]{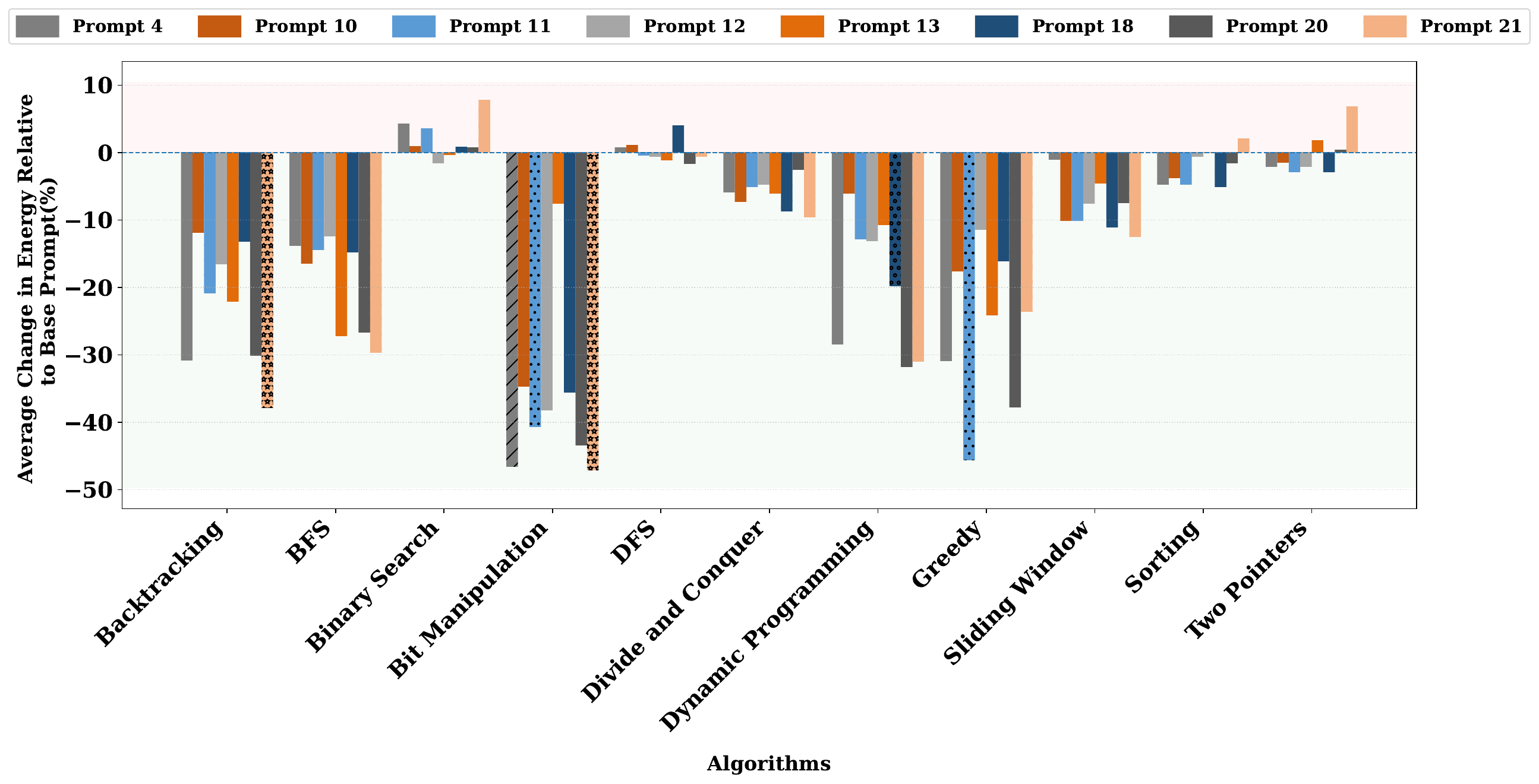}
        }
        \caption{Across algorithmic categories}
        \label{fig:python_algorithm_prompt_impact}
    \end{subfigure}

    \caption{Impact of prompting strategies on the energy consumption of Python code generation relative to the baseline prompt across problem difficulty levels and algorithmic categories, aggregated across all evaluated models.}
    \label{fig:python_prompt_impact}
\end{figure*}

\subsubsection{Overall Impact of Prompts on All the LLMs under Study}

Several models exhibited substantial negative relative changes in energy consumption compared with the baseline prompt. \mbox{Granite-4.0-H-Small} showed a relative change of up to $-50\%$ with Prompt 20, followed by Prompt 11 ($-47\%$) and Prompt 4 ($-42\%$). For \mbox{Claude 4.5 Haiku}, Prompt 21 resulted in a relative change of $-39\%$, followed by Prompt 20 ($-38\%$), Prompt 13 ($-37\%$), and Prompt 11 ($-25\%$). \mbox{MiniMax M3} exhibited relative changes ranging from $-17\%$ to $-28\%$, with Prompt 13 producing the largest decrease ($-28\%$).

The remaining models exhibited comparatively smaller reductions. For \mbox{Gemini 3.8 Flash}, Prompts 20 and 21 each resulted in a $-11\%$ relative change, followed by Prompt 13 at $-10\%$. \mbox{Gemma 4 26B A4B IT} showed a $-8\%$ relative change with Prompts 21 and 13, and $-7\%$ with Prompt 20. For \mbox{GPT-5.6 Terra}, Prompts 4, 12, and 20 each resulted in a $-5\%$ relative change. For GPT-OSS-120B, Prompts 21 and 10 resulted in relative changes of $-11\%$ and $-9\%$, respectively. \mbox{Kimi K2.7 Code} exhibited relative changes of $-8\%$ and $-7\%$ with Prompts 10 and 20, respectively. For \mbox{Muse Glimmer 30B}, Prompts 13, 21, and 20 each resulted in a $-10\%$ relative change. Similarly, Qwen3-Coder-480B-A35B-Instruct exhibited relative changes of $-10\%$, $-8\%$, and $-7\%$ with Prompts 21, 11, and 12, respectively.

Conversely, some prompting strategies resulted in increase in energy consumption compared with the baseline. \mbox{GPT-OSS-120B} exhibited the largest increases with Prompt 13 ($+74\%$) and Prompt 20 ($+33\%$), while Prompt 12 resulted in a $+9\%$ relative change for \mbox{Claude 4.5 Haiku}. Overall, the magnitude and direction of the relative change varied across models and prompting strategies.

The variations in energy consumption across models and prompting strategies described above are illustrated in Figure~\ref{fig:python_energy_reductions}.

\subsubsection{Impact of Prompts on Problem Difficulty}

When examining the overall impact of the prompting strategies across the models under study, we observe overall reductions in energy consumption across all three difficulty categories relative to the baseline prompt. For Hard problems, Prompt 21 and Prompt 12 achieved the largest overall reductions, with both reaching up to $-33\%$. The remaining prompts, with the exception of Prompt 13 ($-5\%$), also resulted in substantial reductions with relative changes ranging from $-28\%$ to $-33\%$. A similar trend was observed for Medium problems, with all prompting strategies achieving overall energy reductions. Prompt 20 yielded the largest reduction at $-28\%$, followed by Prompt 11 ($-23\%$), Prompt 4 ($-22\%$), and Prompt 21 ($-20\%$).In contrast, the impact on Easy problems was more limited, with all prompts producing changes between $-1\%$ and $-3\%$ relative to the baseline. Figure~\ref{fig:python_difficulty_prompt_impact} shows the impact of prompting strategies across problem difficulty levels described above.

\begin{table*}[t]
\centering
\label{tab:cpp_difficulty_algorithm}
\resizebox{\textwidth}{!}{%
\begin{tabular}{lcccccccccccc}
\toprule
& \multicolumn{12}{c}{\textbf{Algorithms}} \\
\cmidrule(lr){2-13}
\textbf{Difficulty} &
\textbf{Backtracking} &
\textbf{BFS} &
\textbf{Binary Search} &
\textbf{Bit Manip.} &
\textbf{DFS} &
\textbf{Divide \& Conquer} &
\textbf{DP} &
\textbf{Greedy} &
\textbf{Sliding Window} &
\textbf{Sorting} &
\textbf{Two Pointers} &
\textbf{Total} \\
\midrule
\textbf{Easy} & 1 & 0 & 20 & 19 & 0 & 1 & 5 & 31 & 4 & 56 & 26 & \cellcolor{gray!30}121 \\
\textbf{Medium} & 9 & 9 & 55 & 40 & 13 & 3 & 105 & 132 & 37 & 97 & 43 & \cellcolor{gray!30}373 \\
\textbf{Hard} & 0 & 7 & 22 & 11 & 2 & 1 & 66 & 15 & 8 & 21 & 1 & \cellcolor{gray!30}113 \\
\textbf{Total} & \cellcolor{gray!10}\textbf{10} & \cellcolor{gray!10}\textbf{16} & \cellcolor{gray!10}\textbf{97} & \cellcolor{gray!10}\textbf{70} & \cellcolor{gray!10}\textbf{15} & \cellcolor{gray!10}\textbf{5} & \cellcolor{gray!10}\textbf{176} & \cellcolor{gray!10}\textbf{178} & \cellcolor{gray!10}\textbf{49} & \cellcolor{gray!10}\textbf{174} & \cellcolor{gray!10}\textbf{70} & \cellcolor{gray!30}\textbf{607} \\

\bottomrule
\end{tabular}
}
\caption{Distribution of benchmark problems solved by all 10 evaluated LLMs across difficulty levels and algorithm categories for C++ code generation.}
\label{tab:cpp_dist}
\end{table*}

\begin{figure*}[t]
    \centering
    \fbox{\includegraphics[width=0.98\textwidth]{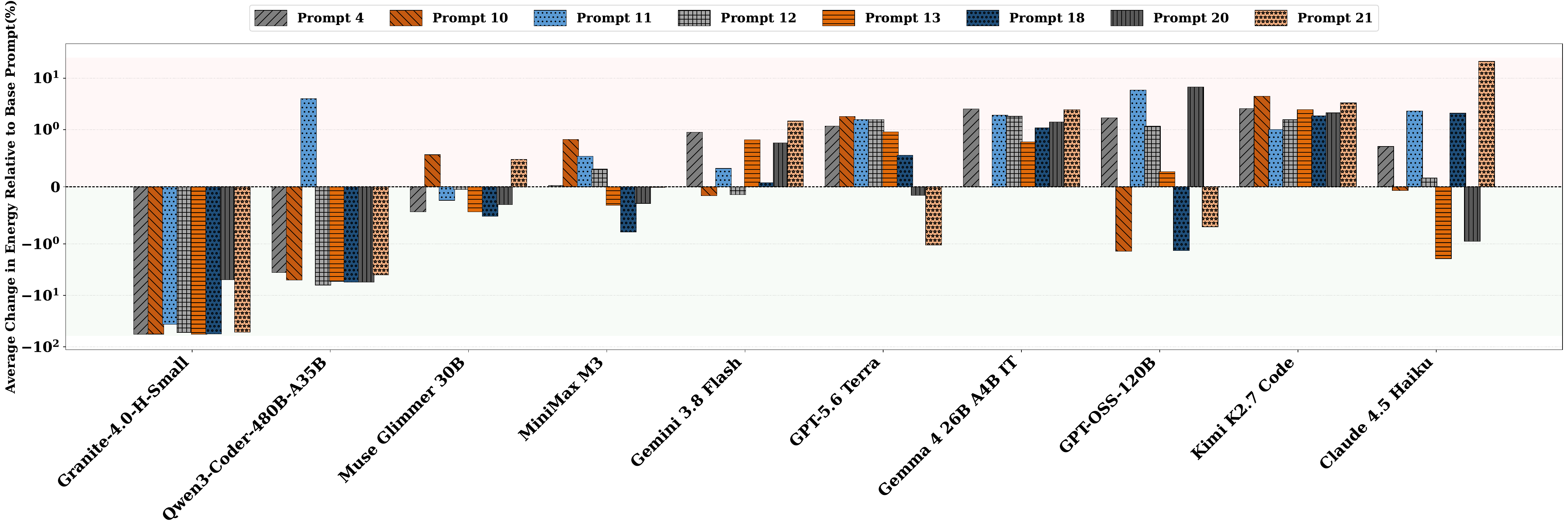}}
    \caption{Average percentage energy reduction achieved by the 8 prompting strategies relative to the baseline prompt for C++ code generation. Negative values indicate lower energy consumption than the baseline.}
    \label{fig:cpp_energy_reductions}
    \vspace{-3mm}
\end{figure*}

Although the aggregate results consistently indicate energy reductions, model-specific outliers were observed across difficulty categories, with some prompting strategies causing a relative increase in energy consumption compared to the baseline. For Hard Problems, \mbox{Claude 4.5 Haiku} showed the largest increases with Prompts 4 and 18, at $+67\%$ and $+11\%$, respectively, while Prompt 20 increased energy consumption by $+22\%$ for \mbox{MiniMax M3}. Qwen3-Coder-480B-A35B-Instruct also exhibited increases with Prompts 4 and 20, at $+21\%$ and $+16\%$. For Medium Problems, Prompt 12 showed a relative increase by $+14\%$ for \mbox{Claude 4.5 Haiku}, while \mbox{GPT-OSS-120B} showed increases of $+48\%$ and $+109\%$ with Prompts 20 and 13, respectively. For Easy Problems, Qwen3-Coder-480B-A35B-Instruct exhibited increases for all prompts except Prompts 11 and 18, with Prompt 4 producing the largest increase at $+2.5\%$. \mbox{GPT-5.6 Terra} showed smaller increases across all prompts, with Prompts 11 and 13 yielding the largest relative changes at $+1.6\%$.

\subsubsection{Impact of Prompts on Algorithmic Category}
When examining the impact of the prompting strategies across models and algorithmic categories, most prompts resulted in lower energy consumption relative to the baseline, although the magnitude of the reduction varied across categories. Several categories showed consistently negative relative changes, with a few prompts producing comparatively larger reductions. For Backtracking, Prompt 21 achieved the largest reduction at $-38\%$, followed by Prompt 4 ($-31\%$) and Prompt 20 ($-30\%$). However, Prompt 20 increased energy consumption by $+43\%$ for \mbox{MiniMax M3}. For BFS, Prompt 21 produced the largest reduction at $-30\%$, followed by Prompts 13 and 20, both at $-27\%$. 

For Bit Manipulation, all prompts resulted in reductions ranging from $-34\%$ to $-47\%$, except Prompt 13, which yielded a smaller reduction of $-7\%$. Prompt 21 achieved the largest reduction at $-47\%$; however, Prompt 20 increased energy consumption for \mbox{MiniMax M3} and \mbox{GPT-OSS-120B} by $+41\%$ and $+11\%$, respectively. For Dynamic Programming, Prompt 20 achieved the largest reduction at $-32\%$, followed by Prompt 21 ($-31\%$) and Prompt 4 ($-28\%$). Despite these overall reductions, Prompt 4 increased energy consumption by $+68\%$ for \mbox{Claude 4.5 Haiku}, while Prompt 11 increased it by $+25\%$ for \mbox{MiniMax M3}.

Greedy exhibited the largest overall reduction among the categories, with Prompt 11 achieving a $-46\%$ reduction, followed by Prompt 20 at $-38\%$. However, Prompt 12 resulted in increased energy consumption for \mbox{Claude 4.5 Haiku} ($+29\%$) and \mbox{Kimi K2.7 Code} ($+21\%$).

More modest but consistent reductions were observed for Divide and Conquer and Sliding Window, with Prompts 21 ($-9\%$) and 18 ($-8\%$) producing the largest reductions for Divide and Conquer, and Prompts 21 ($-12\%$) and 18 ($-11\%$) for Sliding Window.

In contrast, several categories exhibited mixed effects across prompting strategies. For Two Pointers, Prompts 18 and 11 each resulted in a $-3\%$ relative change, whereas Prompt 21 resulted in a $+7\%$ change. Similarly, for Sorting, Prompts 18, 11, and 4 yielded the largest reductions, each with a $-5\%$ relative change, while Prompt 21 resulted in a $+2\%$ change. The effect was comparatively small for DFS, with most prompts producing minimal changes; however, Prompt 18 resulted in a $+4\%$ relative change. Binary Search likewise exhibited relatively limited changes overall, although Prompts 21 and 4 resulted in increases of $+8\%$ and $+4\%$, respectively.

Figure~\ref{fig:python_algorithm_prompt_impact} illustrates the impact of prompting also varies across algorithmic categories.

\subsubsection{Overall Best Performing Prompts for All Models}
Since most of the filtered prompting strategies reduced energy consumption for Python code generation across the evaluated models, we further examined the prompts that consistently improved energy efficiency across all models. Prompts 21 ($-15\%$), 11 ($-12\%$), 18 ($-9\%$), and 10 ($-9\%$) reduced energy consumption relative to baseline across every evaluated model, indicating more consistent energy efficiency improvements across models. Notably, Prompt 21 extends Prompt 10 with an additional energy efficiency prompting step and achieved a larger reduction in energy consumption ($-15\%$ versus $-9\%$). This suggests that the additional prompting step may provide an incremental benefit in improving the energy efficiency of generated code.

\begin{figure*}[t]
    \centering

    \begin{subfigure}[t]{0.48\linewidth}
        \centering
        \fbox{%
            \includegraphics[
                width=0.96\linewidth,
                height=0.50\textheight,
                keepaspectratio
            ]{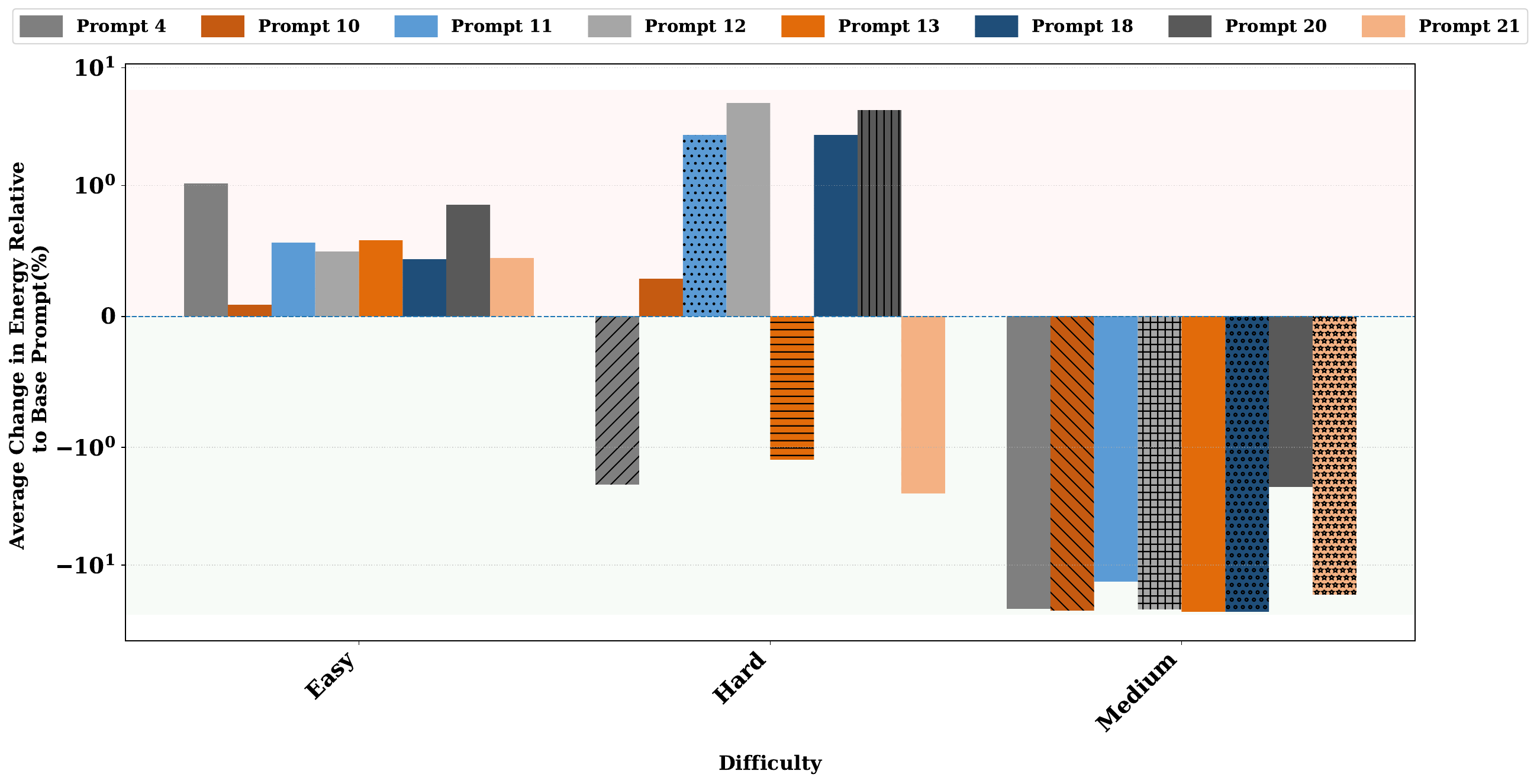}
        }
        \caption{Across problem difficulty levels}
        \label{fig:cpp_difficulty_prompt_impact}
    \end{subfigure}
    \hfill
    \begin{subfigure}[t]{0.48\linewidth}
        \centering
        \fbox{%
            \includegraphics[
                width=0.96\linewidth,
                height=0.50\textheight,
                keepaspectratio
            ]{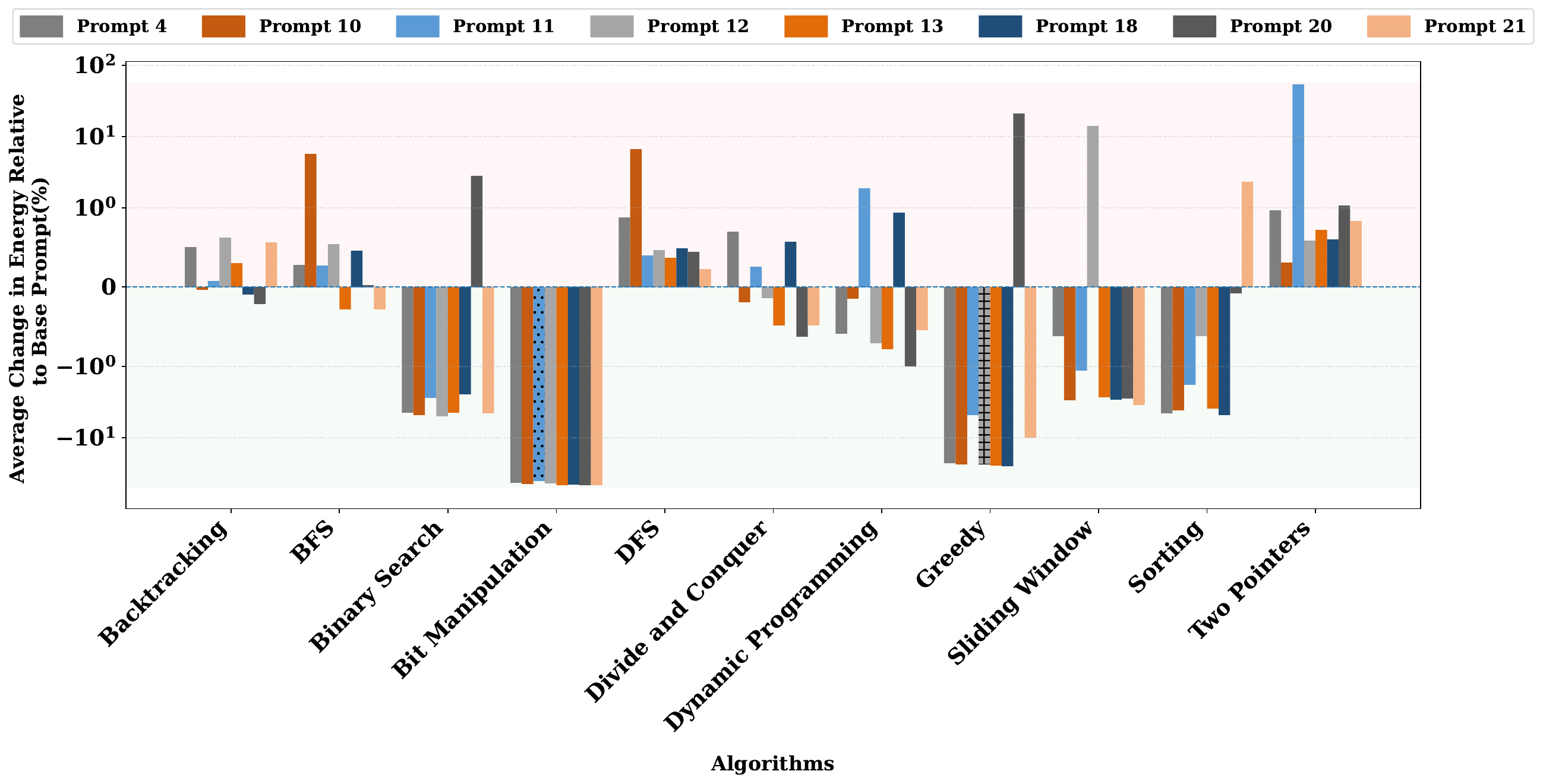}
        }
        \caption{Across algorithmic categories}
        \label{fig:cpp_algorithm_prompt_impact}
    \end{subfigure}

    \caption{Impact of prompting strategies on the energy consumption of C++ code generation relative to the baseline prompt across problem difficulty levels and algorithmic categories, aggregated across all evaluated models.}
    \label{fig:cpp_prompt_impact}
\end{figure*}

\subsection{Evaluation of Prompts for C++ Code Generation}

\subsubsection{Dataset Distribution}
The common C++ evaluation set comprises 607 problems that were solved correctly by all ten evaluated models across every prompting strategy. Among these, 373 problems are classified as Medium difficulty, while 121 are Easy and 113 are Hard. The evaluation set covers a diverse range of algorithmic categories, with the largest representation from Greedy (178), Dynamic Programming (176), and Sorting (174). Other well-represented categories include Binary Search (97), Bit Manipulation (70), and Two Pointers (70). A complete breakdown of the evaluation set by algorithmic category and difficulty level is provided in Table~\ref{tab:cpp_dist}.
\subsubsection{Overall Impact of Prompts on All the LLMs under Study}

The average decrease in the energy consumed by the C++ code generated for each
prompt compared to the baseline is shown in Figure~\ref{fig:cpp_energy_reductions}.

\mbox{Granite-4.0-H-Small} achieved the largest overall reduction in energy consumption, with the largest reduction reaching 57\% across nearly all evaluated prompts, except for Prompt 20, which achieved a reduction of only 4\%. Qwen3-Coder-480B-A35B-Instruct achieved an overall reduction in energy consumption of up to 6.4\%. However, Prompt 11 increased energy consumption by 4\% relative to the baseline.

For \mbox{Muse Glimmer 30B}, \mbox{MiniMax M3}, \mbox{Gemini 3.8 Flash}, and \mbox{GPT-5.6 Terra}, the impact of prompting on energy consumption is relatively small compared with their respective baselines, with energy reductions ranging from $-0.44\%$ to 0.56\%, $-0.79\%$ to 0.83\%, $-0.15\%$ to 1.46\%, and $-1.06\%$ to 1.8\% respectively. These narrow ranges suggest that, for these models, varying the prompting strategy has limited influence on energy consumption relative to the baseline prompt. In contrast to models exhibiting larger variations, their energy consumption appears relatively stable across different prompting strategies.

For \mbox{Gemma 4 26B A4B IT}, \mbox{Kimi K2.7 Code}, GPT-OSS-120B, and \mbox{Claude 4.5 Haiku}, most prompting strategies result in a modest increase in energy consumption relative to their respective baselines, generally within 5\%. However, several prompts produce more substantial increases. For GPT-OSS-120B, Prompts 12 and 20 increase energy consumption by 5.8\% and 6.8\%, respectively, while Prompt 21 for \mbox{Claude 4.5 Haiku} produces a larger increase of 21\%. These results suggest that although prompting generally has a limited effect on energy consumption for these models, certain prompting strategies can introduce a noticeable computational overhead, with the magnitude of this effect varying across models and prompting strategies.

\subsubsection{Impact of Prompts on Problem Difficulty}

In Figure ~\ref{fig:cpp_difficulty_prompt_impact}, the impact of prompting on energy consumption varies substantially across problem difficulty levels. For Medium problems, all evaluated prompts show energy reductions, with the largest reduction reaching 25\% relative to the baseline. However, there were a few model-specific outliers observed for Medium-difficulty problems. For \mbox{Claude 4.5 Haiku}, Prompt 21 resulted in a $+32\%$ increase relative to the baseline prompt, while Prompt 11 exhibited a smaller increase of $+8\%$ for \mbox{Granite-4.0-H-Small}. For Hard problems, the reductions are considerably more limited, with Prompt 4 achieving a 2.05\% reduction, Prompt 13 a 1.27\% reduction, and Prompt 21 a 2.45\% reduction. In contrast, all prompts for Easy problems result in slightly higher energy consumption than the baseline, with the maximum increase up to 1.04\%. Overall, these results indicate that the impact of prompting on energy efficiency is strongly dependent on problem difficulty, with the most substantial energy savings observed for medium-difficulty problems.

\subsubsection{Impact of Prompts on Algorithmic Category}

\begin{figure*}[t]
    \centering

    \begin{subfigure}[t]{0.48\linewidth}
        \centering
        \fbox{%
            \includegraphics[
                width=\linewidth,
                height=50mm
            ]{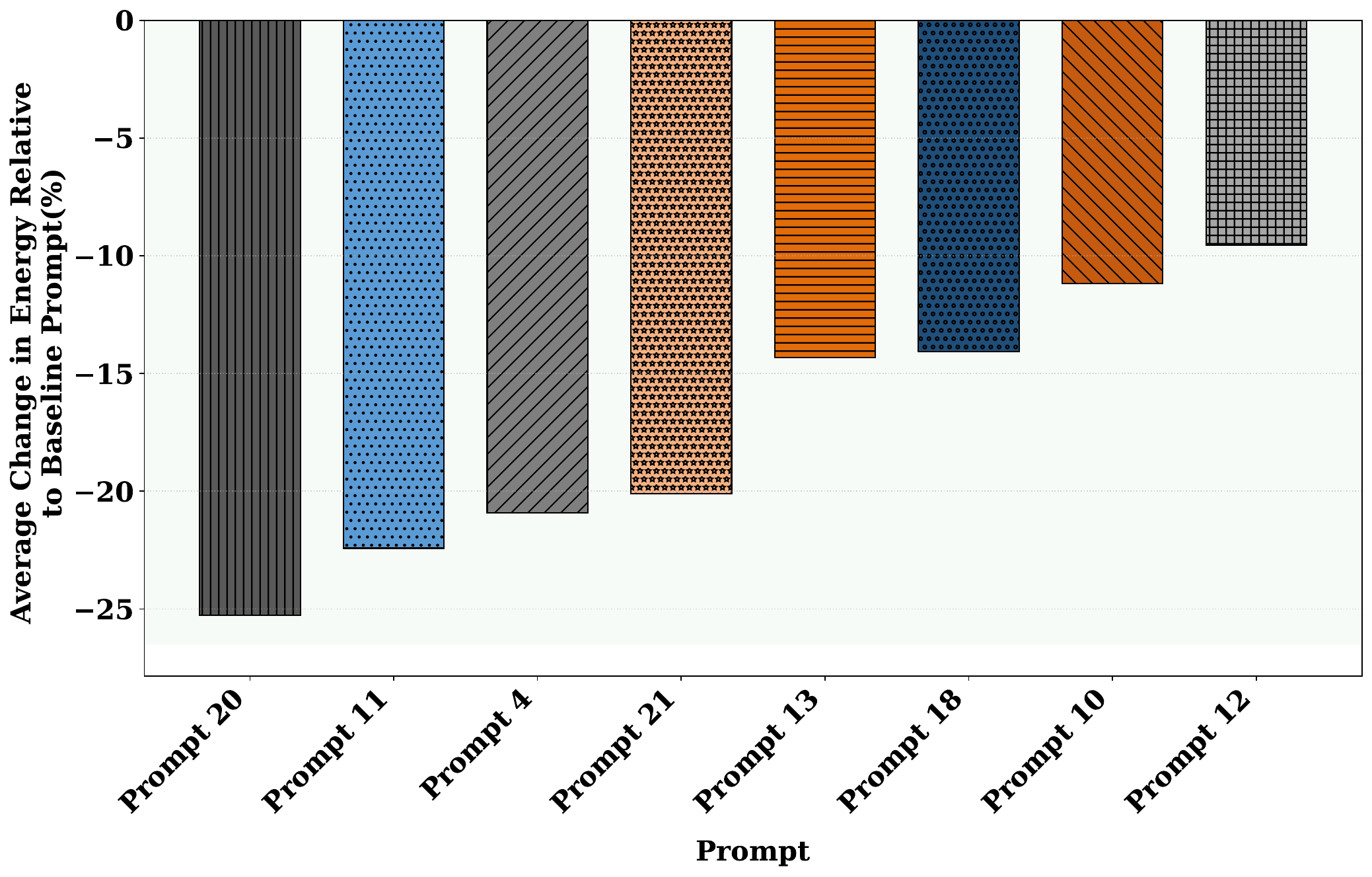}
        }
        \caption{Impact on Python code generation.}
        \label{fig:python_energy_overall_prompts_impact}
    \end{subfigure}
    \hfill
    \begin{subfigure}[t]{0.48\linewidth}
        \centering
        \fbox{%
            \includegraphics[
                width=\linewidth,
                height=50mm]{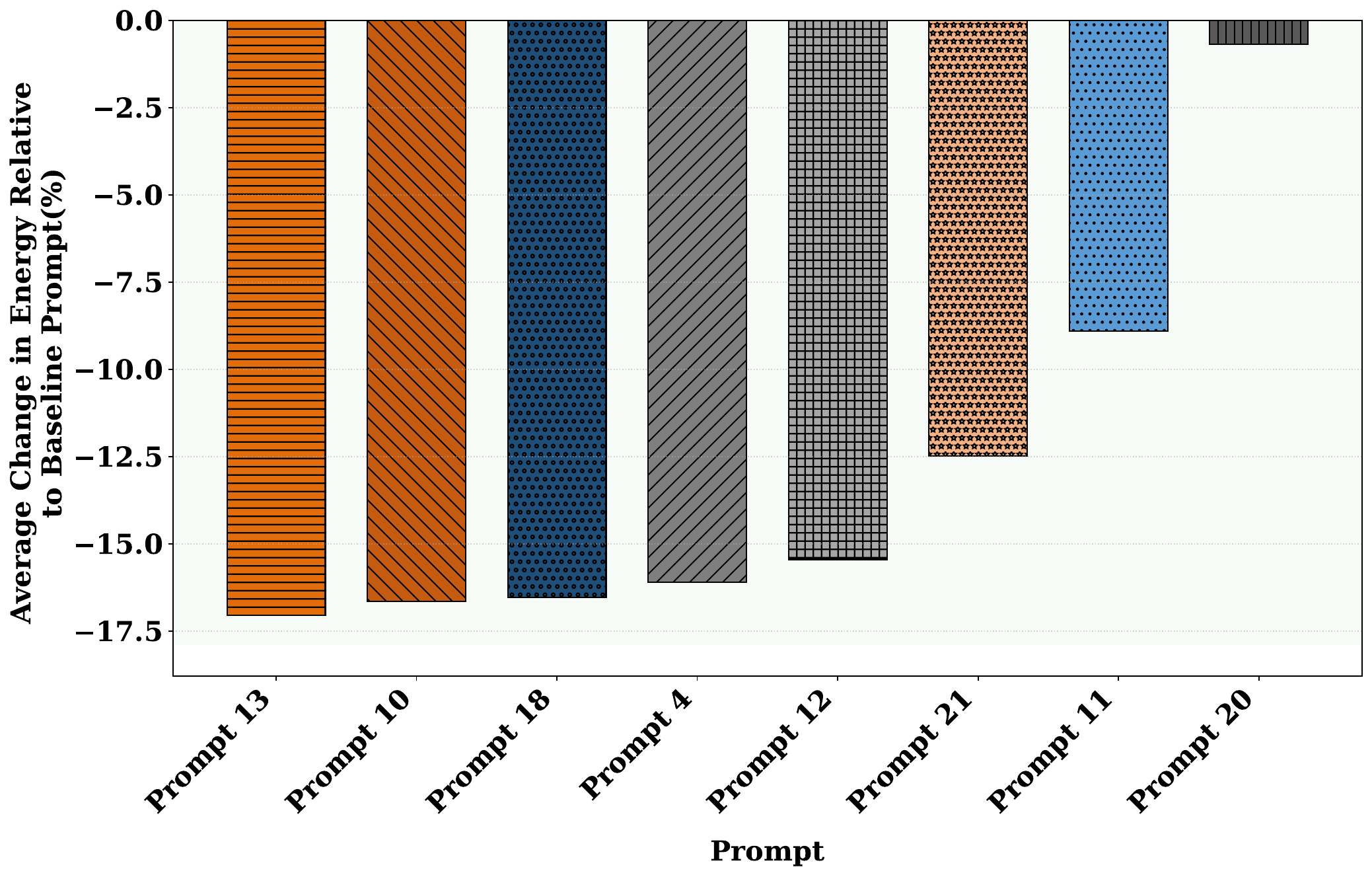}
        }
        \caption{Impact on C++ code generation.}
        \label{fig:cpp_energy_overall_prompts_impact}
    \end{subfigure}

    \caption{Overall impact of 8 prompting strategies Python and C++ code generation for their respective evaluation datasets aggregated across all models.}
    \label{fig:python_cpp_overall_prompts_impact}
\end{figure*}

We further investigate the impact of prompting on energy consumption across different algorithmic categories in C++ code generation. As shown in Figure~\ref{fig:cpp_algorithm_prompt_impact}, prompting is particularly effective for problems in the Bit Manipulation, Greedy, Sliding Window, Sorting, and Binary Search categories, where the evaluated prompts generally achieve more substantial reductions in energy consumption. In the Bit Manipulation category, all evaluated prompts achieved reductions relative to the baseline, with the largest reduction reaching $47\%$. However, a few model-specific outliers deviated from this overall trend. Prompt 11 resulted in increased energy consumption for \mbox{GPT-OSS-120B}, \mbox{Claude 4.5 Haiku}, and \mbox{GPT-5.6 Terra}, with increases of $+31\%$, $+14\%$, and $+6\%$, respectively. For \mbox{Muse Glimmer 30B}, both Prompt 21 and Prompt 10 resulted in a $+9\%$ increases relative to baseline. Prompt 12 also showed increases for \mbox{GPT-OSS-120B} and \mbox{GPT-5.6 Terra} by $+7\%$ and $+5\%$, respectively. Finally, Prompt 4 resulted in increases of $+7\%$ for \mbox{GPT-OSS-120B} and $+5\%$ for \mbox{Gemma 4 26B A4B IT}.

For the Greedy category, all prompts except Prompt 20 resulted in a relative decrease in energy consumption, except Prompt 20 which instead showed $+21\%$ increase. Among the prompting strategies, Prompts 18 ($-25\%$), 13 ($-25\%$), 12 ($-24\%$), 10 ($-24\%$), and 4 ($-23\%$) achieved the largest reductions. Despite this overall trend, there were several model-specific outliers that exhibited relative increases compared to baseline. Prompt 21 resulted in a $+84\%$ increase for \mbox{Claude 4.5 Haiku}. Prompt 10 and Prompt 11 also showed increases for \mbox{Kimi K2.7 Code} and \mbox{GPT-OSS-120B} by $+15\%$ and $+14\%$, respectively. In the Binary Search category, all prompts except Prompt 20 resulted in relative reductions in energy consumption, ranging from $-3\%$ to $-5\%$, while Prompt 20 showed a slight increase of $+3\%$. A model-specific outlier was observed for Prompt 11, which increased energy consumption for \mbox{MiniMax M3} by $+7\%$ compared to baseline.

In the Sorting category, Prompts 4 and 18 resulted in the largest relative reductions of $-5\%$, followed by Prompts 10 and 13, which each achieved a $-4\%$ reduction. Despite this overall trend, Prompt 10 exhibited a model-specific outlier, increasing energy consumption for \mbox{Kimi K2.7 Code} by $+15\%$. In the Sliding Window category, all prompting strategies except Prompt 12 resulted in reductions in energy consumption relative to the baseline. Prompts 10, 13, 18, 20, and 21 achieved the largest relative reduction of $-3\%$, while Prompt 12 exhibited an increase of $+14\%$.

For the remaining categories, the evaluated prompts provide more limited and less consistent energy savings. In particular, all evaluated prompts increase energy consumption for problems in the DFS and Two Pointers categories, with Prompt 11 in the Two Pointers category exhibiting an increase of up to 53.4\% relative to the baseline. Overall, these results indicate that the  impact of prompting on energy efficiency is highly dependent on the underlying algorithmic characteristics of the problem, in addition to the model and prompting strategy.

\subsubsection{Overall Best Performing Prompts for All Models}
Although each model exhibited its own best-performing prompting strategy, Prompts 20 and 18 demonstrated the strongest overall improvements across models. Prompt 20 achieved an overall energy reduction of 2\% across Granite-4.0-H-Small, Qwen3-Coder-480B-A35B-Instruct, \mbox{Muse Glimmer 30B}, \mbox{MiniMax M3}, \mbox{GPT-5.6 Terra}, and \mbox{Claude 4.5 Haiku}, with the largest reduction for an individual model reaching 5\% relative to its corresponding baseline prompt. In comparison, Prompt 18 demonstrated a substantially larger overall reduction of 13\% across \mbox{Granite-4.0-H-Small}, \mbox{Muse Glimmer 30B}, \mbox{MiniMax M3}, Qwen3-Coder-480B-A35B-Instruct and GPT-OSS-120B, with the largest model-level reduction reaching 56\% relative to the baseline prompt.

\subsection{Overall Impact of 8 Energy-Efficient Prompting Strategies For Python and C++ Code Generation}
To summarize the impact of the selected prompting strategies on Python and C++ code generation, we aggregate the energy consumption of code generated using each strategy across the respective datasets and all models under study, and calculate the relative change in energy consumption with respect to the baseline prompt. Overall, the selected prompting strategies resulted in negative relative changes in energy consumption for both Python and C++ code generation, indicating lower energy consumption compared with the baseline.

For Python code generation, Prompt 20 resulted in the largest reduction, with a relative change of $-25\%$, followed by Prompt 11 ($-22\%$), Prompt 4 ($-21\%$), and Prompt 21 ($-20\%$). Prompts 13 and 18 each resulted in a $-14\%$ relative change, while Prompts 10 and 12 resulted in $-11\%$ and $-9\%$, respectively.

For C++ code generation, Prompt 13 resulted in the largest reduction, with a relative change of $-17\%$, followed by Prompts 10, 18, and 4, each with a $-16\%$ relative change. Prompt 12 resulted in $-15\%$, while Prompts 21 and 11 resulted in $-12\%$ and $-9\%$, respectively. In contrast, Prompt 20 resulted in only a $-1\%$ relative change in energy consumption.
The visual summary for the above results can be seen in Figure~\ref{fig:python_cpp_overall_prompts_impact}.

\section{Discussion \& Conclusions}
Explicitly prompting LLMs to prioritize energy efficiency frequently resulted in implementations that consumed less energy than those generated using the baseline prompt, demonstrating that many models can adapt their optimization objective when explicitly instructed. For Python code generation, the largest improvements were observed with Prompt~20 for \mbox{Granite-4.0-H-Small}, reducing energy consumption by approximately $50\%$, Prompt~21 for \mbox{Claude 4.5 Haiku}, achieving a $39\%$ reduction, and Prompt~13 for \mbox{MiniMax M3}, reducing energy consumption by approximately $28\%$. Similar trends were observed for C++ code generation, where \mbox{Granite-4.0-H-Small} achieved a $57\%$ reduction in energy consumption with Prompt~13, while Qwen3-Coder-480B-A35B-Instruct achieved an improvement of $6.4\%$ using Prompt~12. These results suggest that several LLMs already possess the capability to generate more energy-efficient implementations, but this capability is not consistently expressed unless explicitly encouraged through prompting.

Despite these encouraging improvements, the effectiveness of energy-aware prompting varied considerably across models and programming tasks. In several cases, the proposed prompts produced implementations that consumed \emph{more} energy than those generated using the baseline prompt. For instance, Prompt~13 increased the energy consumption of \mbox{GPT-OSS-120B} by up to $74\%$ for Python code generation, while Prompt~21 increased the energy consumption by $22\%$ for C++ code generation. This inconsistency is particularly noteworthy because all programming problems in our benchmark were published by the end of 2023, making it likely that both the benchmark problems and their canonical solutions were included in the training corpora of many evaluated models. Despite this, explicitly instructing models to prioritize energy efficiency did not consistently result in lower energy implementations. Instead, the observed variability suggests that current LLMs do not reliably optimize for energy efficiency during code generation and may instead favor competing objectives such as readability, generality, or commonly learned implementation patterns. Consequently, energy-efficient coding strategies, even if present within the models’ learned representations, are not consistently elicited through prompting alone across different tasks and models.

{\bf Limitations:}
This study has certain limitations:(1) The benchmark used for this evaluation has an imbalanced difficulty distribution, containing substantially more medium-difficulty problems than easy or hard problems.(2) Although the benchmark covers several algorithmic domains, it does not encompass the full diversity of programming tasks encountered in practice. (3) Our experiments were limited to Python and C++ and do not include the effect of our prompting strategies on other programming languages.

{\bf Future Work:}
To address these limitations and the broader challenges of LLM-based code generation, future work could extend this study to additional programming languages to evaluate whether the observed benefits of energy-aware prompting generalize beyond Python and C++. Investigating automated prompt optimization techniques, such as harness engineering, inference-loop optimization, and model-specific prompting strategies tailored to the characteristics of individual LLMs, may further improve the generation of energy-efficient code. Although the selected prompting strategies achieved substantial reductions in execution energy for several models, they also required significantly more tokens, increasing the computational and energy costs of inference. Consequently, the overall energy efficiency of an LLM-based code generation pipeline should account for both the inference energy and the execution energy of the generated code. Future research should therefore explore approaches that improve execution efficiency while minimizing inference overhead, such as fine-tuning LLMs on energy-efficient programming practices or developing concise prompting strategies that consistently elicit energy-efficient implementations.

\section* {Acknowledgments}
This project is in part sponsored by the National Science Foundation (NSF) under award numbers CBET-2343284 and CBET-2447029, by the State University of New York (SUNY), and by IBM Corporation. We also thank Chameleon Cloud for making their resources available for the experiments of this work. We gratefully acknowledge the Fireworks AI team for providing additional computational resources that enabled us to accelerate our experiments.

\small
\bibliographystyle{unsrt}
\bibliography{references}

\end{document}